\RequirePackage[orthodox,l2tabu]{nag}

\documentclass[a4paper,11pt]{article}
\usepackage[T1]{fontenc}        %
\usepackage[utf8]{inputenc}
\usepackage[english]{babel}

\usepackage[left=2.7cm,
            right=2.7cm,
            top=2.7cm,
            bottom=3.2cm
            ]{geometry}        %

\usepackage[protrusion,
            expansion
            ]{microtype} %

\usepackage{booktabs}    %

\usepackage[font=small,labelfont=bf]{caption}
\usepackage{subcaption}

\usepackage{enumitem}

\usepackage{authblk}    %
\usepackage{appendix}
\usepackage{fancyhdr}

\usepackage{braket}
\usepackage{xspace}
\usepackage{mathdots}
\usepackage{stackrel}
\usepackage{xcolor}
\usepackage{mathtools}
\usepackage{graphicx}
\usepackage{amsmath,          %
            amssymb,          %
            amsthm}          %
\usepackage{esint}
\usepackage{xfrac}
\usepackage{comment}

\usepackage{mathrsfs}
\usepackage{bbm}      %
\usepackage{tikz}

\usepackage{slashed}

\usepackage[sort&compress, numbers, square]{natbib}

\definecolor{mycolor1}{HTML}{1F7A8C}
\definecolor{mycolor2}{HTML}{2D3362}
\definecolor{mycolor3}{HTML}{BF1363}
\definecolor{mycolor4}{HTML}{F39237}
\definecolor{mycolor5}{HTML}{2D8B6A}

\usepackage[colorlinks,
            linkcolor=black,
            citecolor=black,
            urlcolor=mycolor3,
            bookmarks,
            bookmarksnumbered
            ]{hyperref}    %

\usepackage{cleveref}
\crefname{figure}{Fig.}{Figs.}
\crefname{equation}{Eq.}{Eqs.}
\crefname{section}{Sec.}{Secs.}

\allowdisplaybreaks

\theoremstyle{definition}

\theoremstyle{remark}

\newcommand{\ii}{\mathrm{i}}

\newcommand{\dd}{\mathrm{d}}

\newcommand{\abs}[1]{\left\vert#1\right\vert}
\newcommand{\avg}[1]{\langle #1 \rangle}

\newcommand{\lr}[1]{\left( {#1} \right)} 
\newcommand{\slr}[1]{\left[{#1} \right]}

\title{\textbf{Entanglement dynamics after quenches with  inhomogeneous Hamiltonians}}

\author[1,2]{Andrea di Pasquale}
\author[1,3,4]{Federico Rottoli}
\author[1,3]{Vincenzo Alba}

\affil[1]{\textit{Dipartimento di Fisica dell’Universit\`a di Pisa,\protect\\Largo B. Pontecorvo 3, I-56127 Pisa, Italy.}}
\affil[2]{\textit{Dipartimento di Fisica e Astronomia ``Galileo Galilei'', Universit\`a di Padova,\protect\\Via Marzolo 8, Padua I-35131, Italy}}
\affil[3]{\textit{INFN Sezione di Pisa, Largo B. Pontecorvo 3, I-56127 Pisa, Italy.}}
\affil[4]{\textit{Istituto di Informatica e Telematica, CNR, Area della Ricerca di Pisa,\protect\\Via G. Moruzzi 1, Pisa I-56127, Italy}}

\date{}

\begin{document}
\maketitle

\begin{abstract}
We investigate entanglement dynamics in bipartite systems governed by inhomogeneous Hamiltonians of the form $H = H_L + H_R$, where $H_{L/R}$ acts only on the left or right region and is homogeneous within each region. Focusing on the XX chain and the transverse-field Ising chain, we derive analytical formulas for the entanglement entropy between the two regions in the hydrodynamic limit of long times. In this regime, fermions incident on the interface undergo scattering, generating entanglement between reflected and transmitted modes. The resulting quasiparticle picture is controlled by the transmission coefficient, which we obtain analytically by solving the stationary lattice Schrödinger equation. Due to the bounded dispersion, strong inhomogeneity suppresses both transport and entanglement growth. We benchmark our analytical predictions against numerical simulations in paradigmatic setups. Finally, we extend the analysis to the interacting XXZ chain using tDMRG. The numerical data show qualitative agreement with the quadratic case: entanglement growth remains suppressed in the strongly inhomogeneous limit. Notably, however, entanglement continues to increase even when transport is suppressed, at least at intermediate times.

\end{abstract}

\tableofcontents
\newpage

\section{Introduction}\label{sec:intro}

The study of out-of-equilibrium quantum many-body systems has proven to be a fruitful research theme in recent years. An important direction aims at characterizing how entanglement is generated in out-of-equilibrium systems, a focus of intense research over the past two decades. Several setups have been investigated, the most prominent being the so-called quantum quench~\cite{polkovnikov2011colloquium}, in which an initial state is evolved under a homogeneous Hamiltonian. An interesting situation is that of bipartite quantum systems, which provide the prototypical setup for studying quantum transport~\cite{alba2022hydrodynamics}. While dynamics from inhomogeneous initial states have been extensively studied~\cite{alba2018entanglementand}, the case of \emph{inhomogeneous} Hamiltonians remains largely unexplored. Interestingly, it has been argued that this setup is relevant for understanding the quantum information paradox in black holes~\cite{kehrein2024page,saha2024generalized,li2025sharp}.

Here we begin exploring entanglement dynamics under inhomogeneous Hamiltonians using the setup depicted in Fig.~\ref{fig:cartoon}(a). We consider a one-dimensional system consisting of two equal-sized parts (left and right in Fig.~\ref{fig:cartoon}). The full system is prepared in an initial state $|\Psi_0\rangle$, which we take to be the product $|\Psi_0\rangle = |\Psi_L\rangle \otimes |\Psi_R\rangle$ obtained by joining two macroscopically different states at $t=0$. For $t>0$, the whole system evolves under an \emph{inhomogeneous} Hamiltonian $H = H_L + H_R$, where $H_L$ ($H_R$) acts nontrivially only on the left (right) half of the system and as the identity elsewhere. We focus on initial states $|\Psi_L\rangle$ and $|\Psi_R\rangle$ that give rise to pairs of entangled quasiparticles under the dynamics generated by $e^{-iH_L}$ and $e^{-iH_R}$, respectively. This implies that deep in the bulk of the two regions, the entanglement dynamics can be understood in terms of the propagation of entangled quasiparticle pairs, in accordance with the standard quasiparticle picture for entanglement spreading in integrable systems~\cite{calabrese-2005,fagotti2008evolution,alba2017entanglement,klobas2021exact,bertini2022growth}. We also consider the case where the right half is prepared in the vacuum state (i.e., the state with no excitations). Specifically, we study dynamics under the anisotropic spin-$1/2$ Heisenberg chain (XXZ chain) with an inhomogeneous magnetic field, described by the Hamiltonian

\begin{equation}
\label{eq:xxz-ham}
    H = - \frac{J}{2} \sum_{j} \slr{\sigma_{j}^{x} \sigma_{j+1}^{x} + \sigma_{j}^{y} \sigma_{j+1}^{y}+\Delta \sigma_j^z\sigma_{j+1}^z} + \sum_{j\leqslant0} h_L  \sigma_{j}^{z} + \sum_{j>0} h_R  \sigma_{j}^{z}\,,
\end{equation}

where $\sigma_j^{x,y,z}$ are Pauli matrices, and $J,\Delta,h_L,h_R$ are real parameters. For $\Delta=0$, Eq.~\eqref{eq:xxz-ham} reduces to the XX chain, which is mappable to free fermions via a Jordan–Wigner transformation. For nonzero $\Delta$ the XXZ chain is interacting but integrable, and its spectrum can be obtained by the Bethe ansatz~\cite{takahashibook}. We also consider the transverse-field Ising chain, defined by

\begin{equation}
\label{eq:tfi-ham}
    H = - J\sum_{j} \sigma_{j}^{x} \sigma_{j+1}^{x} + \sum_{j \leqslant 0} h_L \sigma_{j}^{z} + \sum_{j > 0} h_R \sigma_{j}^{z}\,,
\end{equation}

which can be mapped to a free-fermion model via a Jordan–Wigner transformation and diagonalized by a combination of Fourier transform and Bogoliubov transformation. Our interest is in the dynamics of the entanglement between the left region $A$ and the rest (see Fig.~\ref{fig:cartoon}), quantified by the von Neumann entropy $S = -\mathrm{Tr}\rho_A\ln\rho_A$, where $\rho_A$ is the reduced density matrix of region $A$. We consider the situation where $A$ and its complement are infinite and focus on the long-time limit, i.e., the hydrodynamic regime.

Our main result is that, in the hydrodynamic regime, the entanglement dynamics can be understood within a generalization of the quasiparticle picture~\cite{calabrese-2005,fagotti2008evolution,alba2017entanglement}, as illustrated in Fig.~\ref{fig:cartoon}(b). Let us first focus on models that can be mapped to free-fermion systems. Entangled pairs of quasiparticles are produced in the bulk of the left and right regions. The quasiparticles forming these pairs are the eigenmodes of the left and right Hamiltonians $H_{L/R}$. They travel ballistically with group velocities $v_{L/R}(k)=\varepsilon_{L/R}'(k)$, where $\varepsilon_{L/R}$ are the single-particle energy dispersions of $H_{L/R}$ and the prime denotes derivative with respect to quasimomentum $k$. At the interface between the left and right regions, the dynamics is nontrivial because the eigenmodes of $H_L$ and $H_R$ differ. As shown in Fig.~\ref{fig:cartoon}(b), a left quasiparticle crossing the interface is transformed into an eigenmode of $H_R$ with transmission probability $T(k)$, and is reflected back with probability $R(k)=1-T(k)$. A similar mechanism applies to quasiparticles that are eigenmodes of $H_R$ and approach the interface from the right (not shown in the figure).

Crucially, the transmitted particle is entangled with the reflected one and with the left-moving member of the original pair, thereby generating entanglement between the left and right regions. Notice that the entanglement creation mechanism depicted in Fig.~\ref{fig:cartoon}(b) is similar to that describing entanglement spreading in the presence of localized defects~\cite{eisler2012on,peschel2012exact}. The main difference is that for localized defects, the excitations on the left and right are identical. Our main result is that in systems mappable to free fermions, the entanglement entropy between the two semi-infinite halves grows with time as
\begin{figure}[t]
    \centering
    \includegraphics[width=0.9\linewidth]{figures/cartoon_f.pdf}
    \caption{(a) Cartoon of the setup employed in this work. The system is divided into 
    two parts $L,R$, which are prepared in two different initial states. The system then evolves in time under an \emph{inhomogeneous} Hamiltonian $H=H_L+H_R$, with $H_L$ and $H_R$ acting nontrivially in the left and the right parts of the system, respectively. Here $H_{L/R}$ describe the XXZ chain and the transverse-field Ising chain. We consider initial states that give rise to entangled pairs of quasiparticles in both regions. Entangled quasiparticles travel with opposite velocities, and they are eigenmodes of $H_{L/R}$ (b) Mechanism for entanglement spreading during the dynamics. We focus on the entanglement between a subregion $A$ (the left part of the chain) and the rest. We discuss only the case in which the right region is prepared in the vacuum state at $t=0$. An extensive number of entangled pairs of excitations are produced in the left part of the system only. The right-moving members of the pairs scatter at the interface, giving rise to an entangled triplet formed by the left-moving quasiparticle and the reflected and transmitted quasiparticles. The reflection and transmission probabilities $R,T$ are obtained from the single-particle lattice Schr\"odinger equation and are key ingredients in the quasiparticle picture for entanglement spreading. 
    }
    \label{fig:cartoon}
\end{figure}
\begin{equation}
\label{eq:twoside-intro}
    S =t\int_{0}^{+\pi} \frac{\dd k}{2\pi}\, \abs{v_L(k)} s_{YY}\!\lr{T_{L\to R}\!\lr{k} n_L\!\lr{k}} 
    + t \int_{-\pi}^{0} \frac{\dd k}{2\pi}\, \abs{v_R(k)} s_{YY}\!\lr{T_{R\to L}\!\lr{k} n_R\!\lr{k}}.
\end{equation}

Equation~\eqref{eq:twoside-intro} holds in the hydrodynamic limit as $t\to\infty$. Inspired by the formula for entanglement dynamics in the presence of localized defects, we present it as a conjecture. Here $T_{L\to R}(k)$ is the transmission coefficient for crossing the interface from the left side, $s_{YY}(x)=-x\ln x-(1-x)\ln(1-x)$ is the fermionic Yang–Yang entropy, and $n_{L/R}(k)$ are the momentum occupation densities for the fermions in the left and right parts. In writing Eq.~\eqref{eq:twoside-intro}, we have used the fact that for all models considered, $v_{L/R}(k)=-v_{L/R}(-k)$. Note that $n_L(k)T_{L\to R}(k)$ and $n_R(k)T_{R\to L}(k)$ are the densities of fermions with momentum $k$ that cross the interface from left to right and from right to left, respectively. Equation~\eqref{eq:twoside-intro} admits a simple interpretation: the first and second terms give the contributions of fermions crossing the interface from the left and right sides, respectively. Each fermion crossing the interface is entangled with the quasiparticles that remain on the original side, and its contribution to the entanglement entropy is precisely the Yang–Yang entropy built from the density of crossing fermions. Importantly, because energy is conserved across the interface and the energy dispersions $\varepsilon_{L/R}(k)$ are bounded, transport and entanglement growth are suppressed in the strongly inhomogeneous limit where $H_L$ and $H_R$ differ too much. For instance, for the XX chain, entanglement growth is absent when $|h_L-h_R|>2$. As is clear from Eq.~\eqref{eq:twoside-intro}, the only nontrivial input in the quasiparticle picture is the transmission coefficient; the velocities $v_{L/R}(k)$ and particle densities $n_{L/R}(k)$ are obtained directly from the Hamiltonian and the initial state. In the following, we show that for free-fermion models the transmission coefficients can be derived by solving a single-particle lattice Schr\"odinger equation.

We provide numerical evidence that Eq.~\eqref{eq:twoside-intro} correctly describes the entanglement dynamics after quenches in both the XX chain and the Ising chain (cf.~\eqref{eq:xxz-ham} and~\eqref{eq:tfi-ham}). We should mention that while for the Ising chain Eq.~\eqref{eq:twoside-intro} holds for quenches from generic initial states, this is not the case for the XX chain. Specifically, we observe that Eq.~\eqref{eq:twoside-intro} describes entanglement dynamics for quenches where one of the two parts is prepared in the vacuum state, but it fails if both parts are prepared in nontrivial states. This suggests that the hydrodynamic framework describing transport of quasiparticles (and hence of local conserved quantities) is not sufficient to capture entanglement dynamics. This is not surprising, as the hydrodynamic framework in the presence of defects can be subtle~\cite{takacs2026doubleweaklinkinterferometerhardcorebosons}. It is natural to expect that a quasiparticle picture for entanglement dynamics exists for generic quenches in the XX chain, although its formulation would require a careful treatment of correlations between quasiparticles (see, e.g., Ref.~\cite{caceffo2026fate}). Finally, we also discuss entanglement dynamics in the XXZ chain. Since this model is interacting, we employ the time-dependent density matrix renormalization group~\cite{schollwoeck2011the,paeckel2019time} (tDMRG) to simulate the entanglement dynamics. Similar to the XX chain, in the strongly inhomogeneous regime (i.e., for large $|h_L-h_R|$) the entanglement entropy saturates at long times, whereas it grows linearly with time otherwise.
Surprisingly, our numerical data suggest that, at least for the times accessible with tDMRG, linear entanglement growth persists even when transport of local charges is suppressed.
Although it is challenging to determine whether this is a finite-time effect, this behavior is qualitatively different from that in free systems, where entanglement saturates for $|h_R-h_L|\gtrsim 2$.

The manuscript is organized as follows. In Section~\ref{sec:main} we outline the main result, Eq.~\eqref{eq:twoside-intro}. In Sections~\ref{sec:xx-coeff} and~\ref{sec:ising-coeff} we derive the transmission coefficients for the XX chain and the Ising chain. In Section~\ref{sec:numerics} we provide numerical data supporting Eq.~\eqref{eq:twoside-intro}, with a focus on the XXZ chain in Section~\ref{sec:xxz}. We conclude and discuss future directions in Section~\ref{sec:concl}.

\section{Quasiparticle picture after quenches with inhomogeneous Hamiltonians}\label{sec:theory}
\label{sec:main}

Here we summarize our main result, which is the quasiparticle picture for entanglement spreading after quenches with inhomogeneous \emph{quadratic} Hamiltonians. We discuss the case of fermionic systems, although our results could be extended straightforwardly to bosonic ones. 
Let us consider a generic quadratic free-fermion model, such as the XX chain (see Section~\ref{sec:xx-coeff}) or the Ising chain (see Section~\ref{sec:ising-coeff}). In the thermodynamic limit the model is diagonalized by a combination of Fourier transform and Bogoliubov transformation, and the Hamiltonian can be written as 
\begin{equation}
H=\int \frac{dk}{2\pi}\varepsilon(k) \gamma^\dagger(k)\gamma(k),
\end{equation}
where $\gamma(k)$ are the  Bogoliubov fermion operators diagonalizing the system, $\varepsilon(k)$ the single-particle energy, and the integration is over the momenta $k$ in the first Brillouin zone. 

Let us focus on global quantum quenches~\cite{calabrese2016introduction} from initial states that produce only pairwise entanglement between eigenmodes of the system with opposite momenta (see Fig.~\ref{fig:cartoon} (a)). 
According to the quasiparticle picture~\cite{calabrese-2005,fagotti2008evolution,alba2017entanglement}, in an homogeneous quench the extensive growth of the entanglement entropy between a finite subsystem $A$ of length $\ell$ embedded in an infinite system is given by  
\begin{equation}
\label{eq:quasi}
    S = \int_{-\pi}^{+\pi} \frac{\dd k}{2\pi}\, \min\!\lr{2\abs{v(k)}t, \ell}\, s_{YY}\!\lr{n(k)},
\end{equation}
where $v(k)$ is the group velocity $v(k)=d\varepsilon(k)/dk$ of the quasiparticles, and $n(k)$ is the occupation number of quasiparticle of momentum $k$ in the initial state, i.e., 
\begin{equation}
n(k)=\langle \gamma^\dagger(k)\gamma(k)\rangle, 
\end{equation}
where the expectation is taken over the initial state. Clearly, 
$n(k)$ does not depend on time. Moreover,  $n(k)$ identifies the Generalized Gibbs Ensemble~\cite{vidmar2016generalized} (GGE) that characterizes expectation values of local observables in the steady state. 
In~\eqref{eq:quasi} $s_{YY}$ is the so-called Yang-Yang entropy~\cite{yang1969thermodynamics}, and it
is the thermodynamic entropy of the GGE. The Yang-Yang entropy is written as 
\begin{equation}
\label{eq:yang-yang}
    s_{YY}(n(k)) = - n(k) \ln(n(k)) - \lr{1-n(k)} \ln(1-n(k)),
\end{equation}
The quasiparticle picture interpretation of~\eqref{eq:quasi}  is straightforward. Indeed,  $\min(2|v(k)|t,\ell)$ is the number of entangled pairs formed by the quasiparticles with momenta $\pm k$ that are created at the same point in space at $t=0$ and at a given time $t$ are shared between $A$ and rest. At early times $t\leqslant \ell/(2|v(k)|)$ only pairs created near the edges of subsystem $A$ are shared, and their number grows linearly with time. At long times there is always a quasiparticle in $A$ having an entangled partner outside. This implies that at long times $\min(2|v(k)|t,\ell)=\ell$. As it is clear from~\eqref{eq:quasi}, the entanglement content of the the pairs is the Yang-Yang entropy, which establish an intriguing correspondence between entanglement and thermodynamics. Eq.~\eqref{eq:quasi} holds in the space-time scaling limit in which $\ell,t\to\infty$, with the ratio $t/\ell$ fixed. 
In  the following we focus on the situation in which subsystem $A$ is the half chain. Now,  
Eq.~\eqref{eq:quasi} becomes 
\begin{equation}
    S = t \int_{-\pi}^{+\pi} \frac{\dd k}{2\pi}\, \abs{v(k)} s_{YY}\!\lr{n(k)}, 
\end{equation}
which implies a linear growth at any time. 
Before proceeding we should observe that the fact that the Yang-Yang entropy appears in~\eqref{eq:quasi} is nongeneric. Precisely, this holds only in quenches from initial states that produce entangled \emph{pairs}. The  relationship between the thermodynamic entropy and entanglement breaks down when the initial state produces entanglement between more than two particles, for instance, entangled multiplets~\cite{bertini2018entanglementand,caceffo2023negative}. 

Let us now discuss how the quasiparticle picture outlined so far has to be modified for dynamics with inhomogeneous Hamiltonians. We anticipate that the mechanism for entanglement dynamics is similar to that in free-fermion chains with a ``defect''~\cite{peschel2009reduced} (see also~\cite{ljubotina2019non,alba2021unbounded,capizzi2023domain}). 
The  mechanism is  illustrated in Fig.~\ref{fig:cartoon} (b). Since the Hamiltonian is inhomogeneous, there is an interface at the origin $x=0$. Let us consider the situation in which entangled pairs are produced only in the left part of the chain at $x<0$, whereas the right one is ``empty'', i.e., it does not evolve with time. Now, the right-moving member of each entangled pair undergoes scattering at the interface. Quite generically this leads to a nonzero reflection and transmission amplitudes for the quasiparticle. Importantly, after crossing the interface a quasiparticle produced in the left region gets ``transformed'' into an excitation of $H_R$. 
This is in contrast with dynamics with localized defects, for which the excitations are the same on both sides of the defect. 
As a consequence of the scattering, the transmitted fermion is entangled both with the reflected one and with the left mover of the original pair. 
Inspired by the result for the quasiparticle picture for dynamics in the presence of  defects~\cite{capizzi2023domain}, we conjecture that the dynamics of the entanglement entropy  between the left and right regions is 
\begin{equation}\label{eq:singleside}
    S = t\int_{0}^{+\pi} \frac{\dd k}{2\pi}\, \abs{v(k)} s_{YY}\!\lr{T\!\lr{k} n\!\lr{k}}.  
\end{equation}
Here $T(k)$ is the transmission coefficient, i.e., the probability for a fermion with momentum $k$ originated in the left part to cross the interface. Notice that for generic $H_L,H_R$ $T(k)$ is nonzero only on a subset of $[0,\pi]$, which are the momenta that contribute to the dynamics. In~\eqref{eq:singleside} $s_{YY}(x)$ is the Yang-Yang entropy (cf.~\eqref{eq:yang-yang}), $v(k)=d\varepsilon(k)/dk$ the group velocity, and $n(k)$ the density of Bogoliubov modes in the left part. Notice that in~\eqref{eq:singleside} we exploited the fact that $v(k)=-v(-k)$ to restrict the integration domain in $[0,\pi]$. Notice also that $s_{YY}(T(k)n(k))$ is the Yang-Yang entropy that gets transported in the right region. Clearly, if the Hamiltonian is homogeneous, one has $T(k)=1$, and one recovers the usual quasiparticle picture for quenches from inhomogeneous initial states~\cite{alba2022hydrodynamics}. 
Let us now consider dynamics from initial states that give rise to entangled pairs  in both regions.  
We conjecture that the quasiparticle picture for the entanglement entropy is obtained by summing the contributions of the two regions as 
\begin{equation}\label{eq:twoside}
    S = t\int_{0}^{+\pi} \frac{\dd k}{2\pi}\, \abs{v(k)} s\!\lr{T_{L\to R}\!\lr{k} n_L\!\lr{k}} 
    + t \int_{-\pi}^{0} \frac{\dd k}{2\pi}\, \abs{v(k)} s\!\lr{T_{R\to L}\!\lr{k} n_R\!\lr{k}},
\end{equation}
where $T_{L\to R}$ ($T_{R\to L}$) is the transmission probability from left to right (right to left) and $n_L(k)$ ($n_R(k)$) is the density of quasiparticles generated on the left (right) parts of the chain.
From~\eqref{eq:singleside} and~\eqref{eq:twoside}, it is clear that the transmission coefficients $T_L,T_R$ are the only nontrivial ingredients to render the quasiparticle picture predictive.
In the following sections we derive them for the XX chain and in the Ising chain by solving the lattice single-particle Schr\"odinger equation. 
Finally, we remark that the conjecture in \cref{eq:singleside,eq:twoside} is only valid when the boundary between the left and the right subsystems exactly coincide with the position of the interface. 
In more general scenarios, one would need to consider the triplet between the transmitted, reflected and the original pair, leading to a significantly more complicated expression.

\subsection{Transmission coefficient for the inhomogeneous XX chain}
\label{sec:xx-coeff}

Here we compute the transmission probability of a quasiparticle across the interface in the XX chain. 
After a Jordan-Wigner transformation the XX Hamiltonian in~\eqref{eq:xxz-ham} with $\Delta = 0$ takes the form
\begin{equation}\label{eq:XXhamInhom}
    H =H_L+H_R= -\frac{J}{2}\sum_{j} \slr{c_{j}^\dagger c_{j+1} + c_{j+1}^\dagger c_{j}} + h_L \sum_{j \leqslant 0} c_j^\dagger c_j + h_R \sum_{j > 0} c_j^\dagger c_j  = \frac{1}{2}\sum_{j,l} c_{j}^\dagger \mathcal{H}_{j,l} c_{l},
\end{equation}
where $c_j,c_j^\dagger$ are standard Dirac fermions. 
In~\eqref{eq:XXhamInhom} we defined the single-particle inhomogeneous Hamiltonian $\mathcal{H}_{j,l}$ as  
\begin{equation}
    \mathcal{H}_{j,l} = -\frac{J}{2} \lr{\delta_{j+1,l} + \delta_{j-1,l}} + \begin{cases}
        h_L \delta_{j,l},     &j \leqslant 0,\\
        h_R \delta_{j,l},     &j > 0.\\
    \end{cases}
\end{equation}
The left and right Hamiltonians $H_{L/R}$ in the thermodynamic limit can be diagonalised by going to Fourier space, obtaining
\begin{equation}
    H_{L/R} = \int_{-\pi}^{+\pi}\frac{\dd k}{2\pi}\, \varepsilon_{L/R}(k)\, c_k^\dagger c_k,
\end{equation}
where $c_k=1/L\sum_j e^{-i k j}c_j$ are Fourier transformed fermion operators. Notice that the Fourier modes $c_k$ are the same in the left and right parts of the chain. We anticipate that this is not the case in the Ising chain (see Section~\ref{sec:ising-coeff}). The dispersion relations $\varepsilon_{L/R}(k)$ read as 
\begin{equation}
    \varepsilon_{L/R}(k) = h_{L/R} - J \cos(k). 
\end{equation}
The group velocities $v_{L/R}(k)$ are obtained as 
\begin{equation}
    v_{L/R}(k) = \frac{d\varepsilon_{L/R}(k)}{dk}=J \sin(k), 
\end{equation}
and are the same in the two parts of the system. 

To compute the transmission probability across the interface, we solve the single-particle scattering problem. Precisely, we consider a flux of right moving fermions approaching the interface from the left. 
As ansatz for the scattering eigenfunctions, we consider the superposition of an incoming plane wave from the left with momentum $k_L > 0$ and of a reflected and transmitted waves with unknown amplitudes $\mathcal{R}$ and $\mathcal{T}$. 
Explicitly, the ansatz for the single-particle lattice scattering wavefunction $\Psi_j$ is 
\begin{equation}
\label{eq:ansatz-xx}
    \Psi_j = \begin{cases}
        e^{\ii k_L j} + \mathcal{R}e^{-\ii k_L j},    &j \leqslant 0,\\
        \mathcal{T} e^{\ii k_R j},                    &j > 0.
    \end{cases}
\end{equation}
To proceed, we solve the Schr\"odinger equation $\mathcal{H}_{j,l} \Psi_{l} = E \Psi_{j}\,\forall j$, where we sum over $l$ and we fix the energy $E$. 
Importantly, since the magnetic fields are not the same in the left and right regions, the momentum $k_R >0$ of the transmitted wave will be different from the one of the incoming wave $k_L$. 
In the bulk for $j < 0$ or $j > 1$, the ansatz~\eqref{eq:ansatz-xx} satisfies the Schr\"odinger equation with energy
\begin{equation}\label{eq:xxEnergyCons}
    h_L - J \cos(k_L) = h_R - J \cos(k_R) = E.
\end{equation}
Now, energy conservation~\eqref{eq:xxEnergyCons} can be used to rewrite the momentum $k_R$ of the transmitted wave in terms of  $k_L$. 
We distinguish two different cases. 
If $-1 \leqslant \cos(k_R) = \cos(k_L) + \lr{h_R -h_L}/J \leqslant 1$, Eq.~\eqref{eq:xxEnergyCons} 
admits a real solutions for $k_R$
\begin{equation}\label{eq:xxRightMomentum}
    k_R = \arccos\!\slr{\cos({k_L}) + \frac{h_R}{J} - \frac{h_L}{J}}, 
\end{equation}
which means that transmission through the interface can occur. 
On the other hand, if $\cos(k_L) + \lr{h_R -h_L}/J < -1$ or $\cos(k_L) + \lr{h_R -h_L}/J > 1$ the equation~\eqref{eq:xxEnergyCons} has no real solutions for $k_R$. 
Physically, this corresponds to the fact that there is no asymptotic wave on the right with the 
same energy as the incoming one. 
In this case, the momentum $k_L$ is imaginary, corresponding to a evanescent wave, and the incoming quasiparticle undergoes total reflection. 
For $\abs{h_R -h_L} \geqslant 2$ no modes can be transmitted through the barrier. This implies that the transmission coefficient vanishes for all momenta and (cf.~\eqref{eq:twoside}) the linear growth of the entanglement entropy is absent. 
In the following we restrict ourselves to the situation with $1 \leqslant \cos(k_R) = \cos(k_L) + \lr{h_R -h_L}/J \leqslant 1$. 
To compute the reflection $\mathcal{R}$ and transmission amplitudes $\mathcal{T}$, we need to impose the Schr\"odinger equation at the interface at $j = 0, 1$. After some algebra this yields the boundary conditions
\begin{gather}\label{eq:xxBoundaryEqs}
    1 + \mathcal{R} = \mathcal{T},\\
    \label{eq:xxBoundaryEqs1}
    e^{\ii k_L} + \mathcal{R} e^{-\ii k_L} = \mathcal{T} e^{\ii k_R},
\end{gather}
where $k_R$ is given in \cref{eq:xxRightMomentum}. 
The solution of \cref{eq:xxBoundaryEqs} and~\eqref{eq:xxBoundaryEqs1} are then
\begin{gather}
    \mathcal{R} = \frac{e^{\ii k_L} - e^{\ii k_R}}{e^{\ii k_R} - e^{-\ii k_L}} = \frac{\sin (k_L) - \sin( k_R) +\ii \lr{h_R - h_L}/J}{\sin( k_L) + \sin( k_R) -\ii \lr{h_R - h_L}/J},\\
    \mathcal{T} = \frac{e^{\ii k_L} - e^{-\ii k_L}}{e^{\ii k_R} - e^{-\ii k_L}} = \frac{2 \sin (k_L)}{\sin (k_L) + \sin (k_R) -\ii \lr{h_R - h_L}/J},
\end{gather}
where we have used Eq.~$\eqref{eq:xxEnergyCons}$ to rewrite the expression in terms of $\sin(k_L)$ and $\sin(k_R)$.

Finally we need to compute the reflection $R$ and transmission $T$ probabilities in terms of the amplitudes $\mathcal{R}$ and $\mathcal{T}$.
The relation between the reflection and transmission probabilities and the corresponding amplitudes can be obtained from the continuity equation of the probability current derived from the Schr\"odinger equation.
Using the fact that the incoming, reflected and transmitted waves are plane waves, their (unnormalised) probability currents are simply given by their squared amplitudes times their group velocity
\begin{equation}
    j_0 = v_L(k_L),\qquad j_R = v_L(k_L) \left | \mathcal{R} \right |^2, \qquad j_T = v_R(k_R) \left | \mathcal{T} \right |^2\,,
\end{equation}
where $j_0$ is the probability current of the incoming wave, $j_R$ of the reflected and $j_T$ of the transmitted one. 
The continuity equation then reduces to $j_0 = j_R + j_T$, which implies that the reflection probability is simply the absolute value squared of the reflection amplitude $\mathcal{R}$
\begin{equation}\label{eq:xxR}
    R = \frac{j_R}{j_0} = \abs{\mathcal{R}}^2 = \frac{\lr{\sin(k_L) - \sin(k_R)}^2 + \lr{h_R -h_L}^2/J^2}{\lr{\sin(k_L) + \sin(k_R)}^2 + \lr{h_R -h_L}^2/J^2},
\end{equation}
while the transmission probability has an additional factor coming from the ratio of the different velocities to the left and to the right of the interface
\begin{equation}\label{eq:xxT}
    T = \frac{j_T}{j_0} = \frac{v_R\!\lr{k_R}}{v_L\!\lr{k_L}} \abs{\mathcal{T}}^2 = \frac{4 \sin(k_L) \sin(k_R)}{\lr{\sin(k_L) + \sin(k_R)}^2 + \lr{h_R -h_L}^2/J^2}. 
\end{equation}
Clearly, one has  $R+T = 1$. It is useful to discuss the continuum limit. Let us rewrite $J=1/(ma^2)$, $h=1/(m a^2)+V$, with $a$ the lattice spacing and $m$ the mass of the particle. The continuum limit $a\to0$ gives the single-particle dispersion and the group velocity as 
\begin{equation}
\varepsilon(k)=\frac{k^2}{2m}+V,\quad v(k)=\frac{k}{m}. 
\end{equation}
Finally, from~\eqref{eq:xxR} and~\eqref{eq:xxT} one obtains 
\begin{equation}
R=\frac{(k_L-k_R)^2}{(k_L+k_R)^2},\quad 
T=\frac{4k_L k_R}{(k_L+k_R)^2}, 
\end{equation}
with transmitted momentum
\begin{equation}
    k_R = \sqrt{k_L^2 + 2m \lr{V_L-V_R}},
\end{equation}
which yield the transmission and reflection coefficients for a quantum particle of mass $m$ scattering on a step potential.

\subsection{Transmission coefficient for the inhomogeneous Ising chain}
\label{sec:ising-coeff}

Let us now derive the transmission coefficient in the transverse-field Ising chain with inhomogeneous magnetic field.  
Let us first review the diagonalisation of the homogeneous Ising Hamiltonian.
The Hamiltonian takes the form
\begin{equation}
\label{eq:is-ham}
    H = - \sum_{j = -\infty}^{+\infty} \slr{ t \sigma^x_j \sigma_{j+1}^x - \frac{h}{2}\sigma_j^z }.
\end{equation}
To diagonalise~\eqref{eq:is-ham}, we first map it to a fermionic chain using a Jordan-Wigner transformation, which yields 
\begin{equation}
\label{eq:is-ham-fer}
    H 
    =  -t \sum_{j} \lr{c_j^\dagger c_{j+1} + c_{j+1}^\dagger c_j + c_{j+1} c_{j} + c_{j}^\dagger c_{j+1}^\dagger} -  h \sum_{j} \lr{ c_{j}^\dagger c_{j} - \frac{1}{2}} .
\end{equation}
To proceed, let us introduce the Majorana fermions $a_j,b_j$ as  $c_j^\dagger = \lr{a_j + \ii\, b_j} /\sqrt{2}$. 
This allows us to rewrite~\eqref{eq:is-ham-fer} as 
\begin{equation}
    H =  \frac{\ii}{2}  \sum_j \slr{2 t a_{j} b_{j-1} - 2 t b_{j} a_{j+1} + h a_j b_j - h b_j a_j }
    = \frac{1}{2} \sum_{j,l} \begin{pmatrix}
        a_j &b_j
    \end{pmatrix} \mathcal{H}_{jl}\begin{pmatrix}
        a_l\\
        b_l
    \end{pmatrix}.
\end{equation}
Here we have introduced the single-particle Hamiltonian $\mathcal{H}_{jl}$ as 
\begin{equation}
    \mathcal{H}_{jl} = \ii  \slr{\begin{pmatrix}
        0   &2t\\
        0   &0\\
    \end{pmatrix}\delta_{j-1,l} + \begin{pmatrix}
        0   &0\\
        -2t  &0
    \end{pmatrix}\delta_{j+1,l} + \begin{pmatrix}
        0   &h\\
        -h  &0
    \end{pmatrix}\delta_{j,l}},
\end{equation}
which in Fourier space becomes 
\begin{equation}
    \mathcal{H}(k) = \ii \begin{pmatrix}
        0               &h+2te^{-\ii k}\\
        -h-2te^{\ii k}     &0
    \end{pmatrix} 
    = \varepsilon(k) \begin{pmatrix}
        0                       &-\ii e^{-\ii\theta(k)}\\
        \ii e^{\ii\theta(k)}    &0
    \end{pmatrix}.
\end{equation}
Finally, the eigenvectors of the two-by-two matrix $\mathcal{H}(k)$ are
\begin{equation}
    \Phi_k^\pm = \frac{1}{\sqrt{2}}\begin{pmatrix}
        1\\
        \pm \ii e^{\ii \theta(k)}
    \end{pmatrix},
\end{equation}
where $\theta(k)$ is the so-called Bogoliubov angle
\begin{equation}
    e^{\ii\theta(k)} = -\frac{h + 2 t e^{\ii k}}{\sqrt{4t^2+h^2 + 4th\cos(k)}}. 
\end{equation}
The eigenvalues of $\mathcal{H}(k)$ are $\pm\varepsilon(k)$, giving the  dispersion relation of the model as 
\begin{equation}
\label{eq:disp-is}
    \varepsilon(k) = \sqrt{4t^2 + h^2 + 4th \cos(k)}.
\end{equation}
From the dispersion relation, we find the group velocity of the excitations as  
\begin{equation}
\label{eq:is-vel}
    v(k) = \varepsilon'(k) = -\frac{4t h \sin k}{\varepsilon(k)}.
\end{equation}

Here we consider the inhomogeneous version of the Ising chain, described by the Hamiltonian 
\begin{equation}
    H =  \ii\, t \sum_j \lr{ a_{j} b_{j-1} - b_{j} a_{j+1}} + \frac{\ii}{2}\,h_L \sum_{j\leqslant 0} \lr{a_j b_j - b_j a_j }  + \frac{\ii}{2}\,h_R \sum_{j > 0} \lr{a_j b_j - b_j a_j },
\end{equation}
where we employed the Majorana representation, and $h_L,h_R$ are the magnetic fields in the two halves of the system. 
Similar to the inhomogeneous XX chain, to determine the transmission and reflection probability we employ the ansatz as 
\begin{equation}\begin{split}
    \Psi_j^\pm &= 
     \begin{cases}
        \begin{pmatrix}
            1\\
            \pm \ii e^{\ii \theta_L}
        \end{pmatrix} e^{\ii k_L j} + \mathcal{R}\begin{pmatrix}
            1\\
            \pm\ii e^{-\ii \theta_L}
        \end{pmatrix} e^{-\ii k_L j},  &j\leqslant 0,\\
        \mathcal{T}\begin{pmatrix}
            1\\
            \pm \ii e^{\ii \theta_R}
        \end{pmatrix} e^{\ii k_R j},   &j>0,
    \end{cases}
\end{split}\end{equation}
where to lighten the notation we defined the angles $\theta_L = \theta(k_L)$, $\theta_R = \theta(k_R)$, and  we exploited the explicit form of the Bogoliubov angle to write
\begin{equation}
\label{eq:bogo-is}
    e^{\ii\theta(-k)} = -\frac{h+2te^{-\ii k}}{\varepsilon(k)} = e^{-\ii \theta(k)}.
\end{equation}
The Schr\"odinger equation in the bulk  for $j<0$ and $j>1$ is  satified by imposing energy conservation, which allows to obtain a relationship between $k_R$ and $k_L$ as 
\begin{equation}\label{eq:isingEnergyCons}
    \sqrt{4t^2+ h_L^2 +4 t h_L \cos(k_L)} = \sqrt{4t^2+h_R^2 + 4t h_R \cos(k_R)} = E.
\end{equation}
As was the case for the XX chain, \cref{eq:isingEnergyCons} only admits real solutions for $k_R$ if 
$-1\leqslant \lr{h_L^2 - h_R^2 + 4th_L \cos(k_L)}/\lr{4th_R} \leqslant 1$, in which case the wave is partially transmitted with momentum
\begin{equation}
\label{eq:isi-rela}
    k_R = \arccos\!\slr{\frac{1}{4 t h_R} \lr{h_L^2 - h_R^2 + 4t h_L \cos(k_L)}}.
\end{equation}
After imposing the Schr\"odinger equation at the boundary $j = 0, 1$ we obtain a system of four equations as 
\begin{equation}
    2 \ii J\slr{\begin{pmatrix}
        \pm \ii \lr{ e^{\ii \theta_L} e^{-\ii k_L} + \mathcal{R} e^{-\ii \theta_L} e^{\ii k_L} - h_L e^{\ii\theta_L} - h_L \mathcal{R} e^{-\ii \theta_L}} \\
        - \mathcal{T} e^{\ii k_R} + h_L (1+\mathcal{R})
    \end{pmatrix}} = \varepsilon(k_L) \begin{pmatrix}
        1\\
        \pm \ii e^{\ii \theta_L}
    \end{pmatrix}(1+\mathcal{R}),
\end{equation}
\begin{equation}
    2\ii J \slr{\begin{pmatrix}
        \pm \ii \lr{e^{\ii \theta_L} + \mathcal{R} e^{-\ii \theta_L} - h_R \mathcal{T}  e^{\ii \theta_R} e^{\ii k_R}}\\
        -\mathcal{T} e^{\ii 2 k_R} + h_R \mathcal{T} e^{\ii k_R}
    \end{pmatrix}} = \varepsilon(k_L) \begin{pmatrix}
        1\\
        \pm \ii e^{\ii \theta_R}
    \end{pmatrix} \mathcal{T} e^{\ii k_R}.
\end{equation}
Interestingly, only two of the equations are independent
\begin{gather}
    \mathcal{T} e^{\ii k_R} = e^{\ii k_L} + \mathcal{R} e^{-\ii k_L},\\
    e^{\ii\theta_L}  + \mathcal{R} e^{-\ii\theta_L} = \mathcal{T} e^{\ii \theta_R}.
\end{gather}
which allows us to obtain the reflection and transmission amplitudes as 
\begin{gather}
    \mathcal{R} = 
    \frac{h_R e^{\ii k_L} - h_L e^{\ii k_R}}{h_L e^{\ii k_R} - h_R e^{-\ii k_L}},\\
    \mathcal{T} = 
    \frac{2\ii h_L \sin k_L}{h_L e^{\ii k_R} - h_R e^{-\ii k_L}}. 
\end{gather}
Finally, the transmission and reflection coefficients are obtained as 
\begin{gather}
    R = \abs{\mathcal{R}}^2 = \frac{h_L^2 + h_R^2 -2 h_L h_R \cos\!\lr{k_L - k_R}}{h_L^2 + h_R^2 -2 h_L h_R \cos\!\lr{k_L + k_R}},\\
    \label{eq:Tisi}
    T = \frac{v_L\!\lr{k_L}}{v_R\!\lr{k_R}} \abs{\mathcal{T}}^2 = \frac{4 h_L h_R \sin k_L \sin k_R}{h_L^2 + h_R^2 -2 h_L h_R \cos\!\lr{k_L + k_R}},
\end{gather}
which can be shown to satisfy $T + R = 1$.
Again, let us consider the continuum limit. 
We  rewrite 
$\abs{h+2t} = mc^2$ and  
$t = c/\lr{2a}$ where $c$ is the velocity of sound and $a$ is the lattice spacing, and we take the limit $a \to 0$. 
In this limit the dispersion relation~\eqref{eq:disp-is} yields the correct relativistic dispersion relation 
\begin{equation}
    \varepsilon(k) 
    =\sqrt{m^2 c^4 + k^2 c^2} + \mathcal{O}\!\lr{a}.
\end{equation}
The group velocity of the excitations~\eqref{eq:is-vel} become 
\begin{equation}
    v(k) = 
\frac{kc^2}{\sqrt{m^2 c^4 + k^2c^2}}+\mathcal{O}(a). 
\end{equation}
The Bogoliubov angle $\theta(k)$ (cf.~\eqref{eq:bogo-is}) is given as 
\begin{equation}
    e^{\ii \theta(k)} = 
    -\frac{mc + \ii k}{\sqrt{m^2c^2 + k^2}}+{\mathcal{O}(a)}.
\end{equation}
The relationship between $k_L$ and $k_R$ (cf.~\eqref{eq:isi-rela}) becomes 
\begin{equation}
    k_R %
    =\sqrt{k_L^2 + \lr{m_L^2-m_R^2} c^2} + \mathcal{O}(a),
\end{equation}
which is the result for the relativistic dispersion relation.
The reflection and transmission coefficients  are then 
\begin{equation}
    R = \frac{\lr{k_L -k_R}^2 + \lr{m_L-m_R}^2 c^2}{\lr{k_L + k_R}^2 + \lr{m_L-m_R}^2 c^2},
\end{equation}
\begin{equation}\begin{split}
    T = \frac{4k_L k_R}{\lr{k_L + k_R}^2 + \lr{m_L-m_R}^2 c^2},
\end{split}\end{equation}
and as expected one has $T+R = 1$.

\begin{figure}[t]
    \centering
    \includegraphics[width=0.48\linewidth]{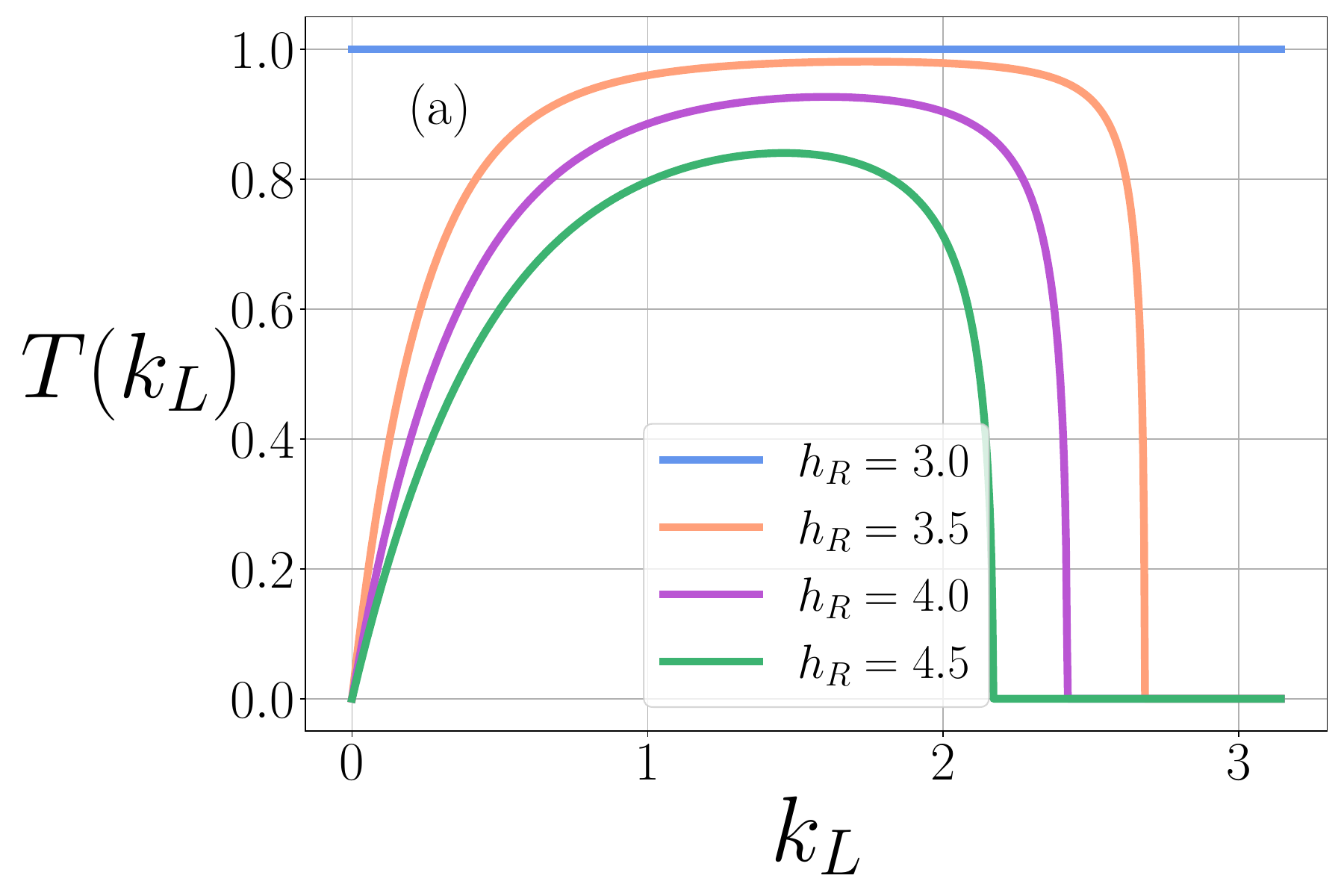}
    \includegraphics[width=0.48\linewidth]{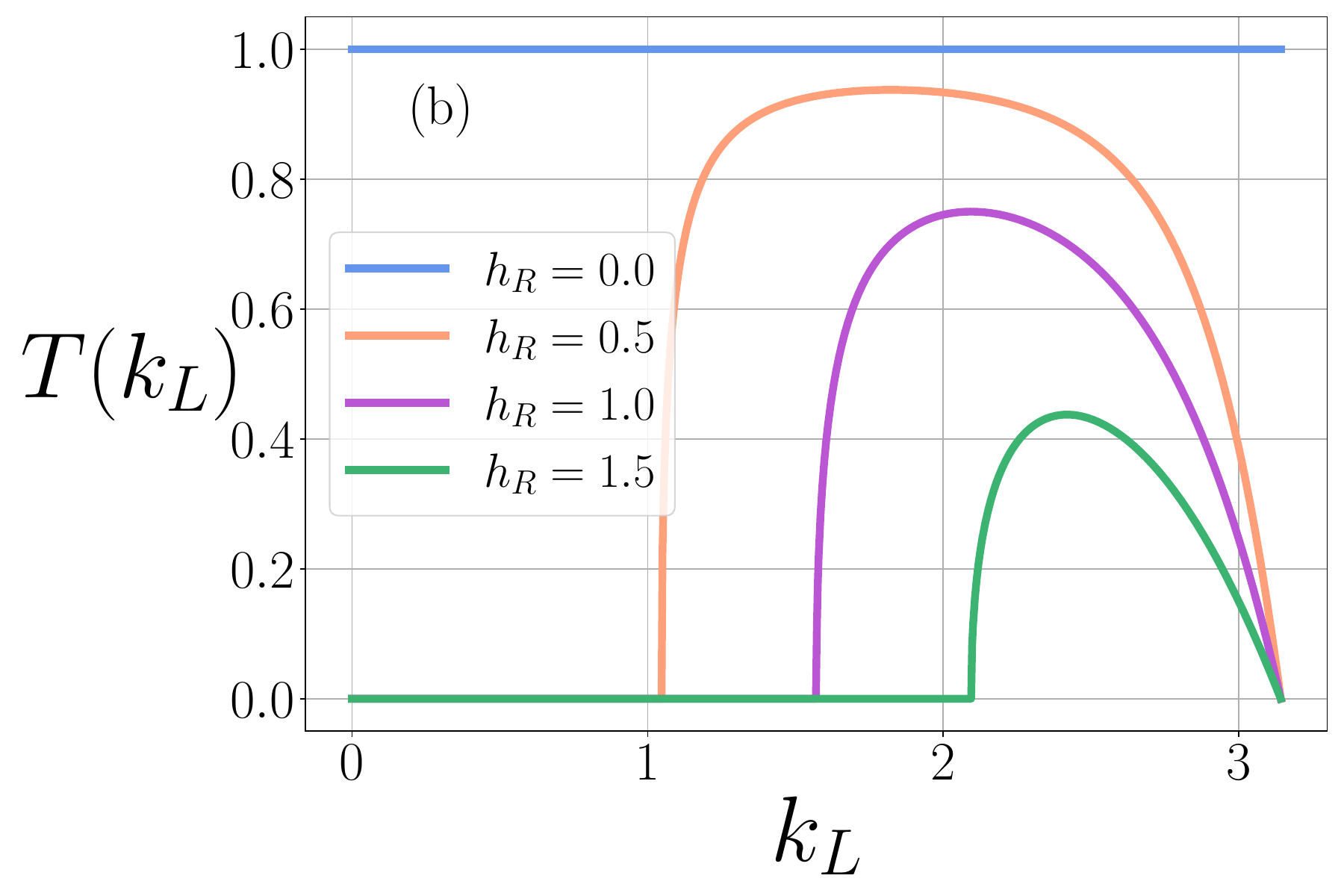}
    \caption{Transmission coefficients $T(k_L)$ for the  inhomogeneous Ising and XX chains (left and right panels, respectively) as a function of the  momentum $k_L$. 
    We consider the setup of Fig.~\ref{fig:cartoon} in which a right-moving particle originated in the left part with momentum $k_L$ is scattering at the interface. 
    We remind that since the momentum $k_R$ on the right of the scattering interface is a function of the one on the left $k_L$, the transmission coefficient is only function of $k_L$.
    For the Ising chain (left panel) we consider the situation in which the magnetic field in the left region is fixed $h_L=3$  and we vary $h_R$. For the $XX$ chain (right panel) we fix $h_L=0$. For $h_R=h_L$ one has $T(k_L)=1$ for any $k_L$. We restrict to $h_L,h_R>0$. Upon increasing $|h_R-h_L|$ the region of quasimomenta for which transmission is possible shrinks; in particular if the bands do not overlap, i.e., $\max_k \varepsilon_L(k) < \min_k \varepsilon_R(k)$, transmission cannot occur.
    }
    \label{fig:transmission}
\end{figure}

In Fig.~\ref{fig:transmission} we show the transmission coefficient $T(k_L)$ for both the Ising chain (panel (a)) and the XX chain (panel (b)). We consider the situation illustrated in Fig.~\ref{fig:cartoon} in which a quasiparticles produced in the left region scatters at the interface. We plot $T(k_L)$ versus $k_L$. 
Recall that, while in the analytic expression for the transmission coefficient it appears both the momentum $k_L$ on the left of the interface and the one $k_R$ on the right, the momentum on the right is fixed by energy conservation and is a function of the one on the left $k_R = k_R(k_L)$, therefore the transmission coefficient is only function of $k_L$, $T = T(k_L)$.
In both panels we fix the value of $h_L$ and show results for several values of $h_R$. Precisely, for the XX chain we consider $h_L=0$, whereas for the Ising chain we choose $h_L=3$.  For $h_L=h_R$ the Hamiltonian is homogeneous, and $T(k_L)=1$ for any $k_L$. For $h_R\ne h_R$ only fermions in a restricted region in momentum space can be transmitted. Upon increasing the difference between $h_L$ and $h_R$ the permitted regions for transport shrink, and eventually vanish when the energy dispersions in the left and right region do not overlap.

\section{Numerical benchmarks}\label{sec:numerics}

In this section, we provide numerical evidence supporting the validity of~\eqref{eq:twoside} in the XX chain and in the Ising chain. To render Eq.~\eqref{eq:twoside} predictive we employ the results in Eq.~\eqref{eq:xxT} and Eq.~\eqref{eq:Tisi} for the transmission coefficient. To go beyond free-fermion models, we also consider the XXZ chain, discussing numerical data obtained by using the time-dependent Density Matrix Renormalization Group~\cite{schollwoeck2011the,paeckel2019time} (tDMRG). %

Before proceeding, let us observe that the dynamics of $S$  can be obtained from the time-evolved two-point Majorana correlation function $\Gamma$ defined as 
\begin{equation}
\Gamma_{jl}=\left(
\begin{array}{cc}
\langle a_ja_l\rangle & \langle a_jb_l\rangle\\
\langle b_j a_l\rangle & \langle b_jb_l\rangle
\end{array}\right),
\end{equation}
where $a_j,b_j$ are time-dependent Majorana operators (see Section~\ref{sec:ising-coeff}) and the 
expectation value is taken with respect to the initial state. The dynamics of $\Gamma$ is obtained by solving the Heisenberg equations 
\begin{equation}
\label{eq:heis}
\frac{d \Gamma(t)}{dt}=\frac{i}{\hbar}[\mathcal{H},\Gamma(t)],
\end{equation}
with $\mathcal{H}$ the inhomogeneous Ising or XX Hamiltonians in the Majorana basis. Since the matrix $\Gamma$ is $2L\times 2L$, the computational cost to solve~\eqref{eq:heis} grows as $L^3$. Since we choose Gaussian initial states, and the dynamics preserves Gaussianity, the entanglement entropy is straightforwardly obtained from $\Gamma$ restricted to region $A$~\cite{peschel2009reduced}. 

\begin{figure}[t]
    \centering
    \includegraphics[width=0.7\linewidth]{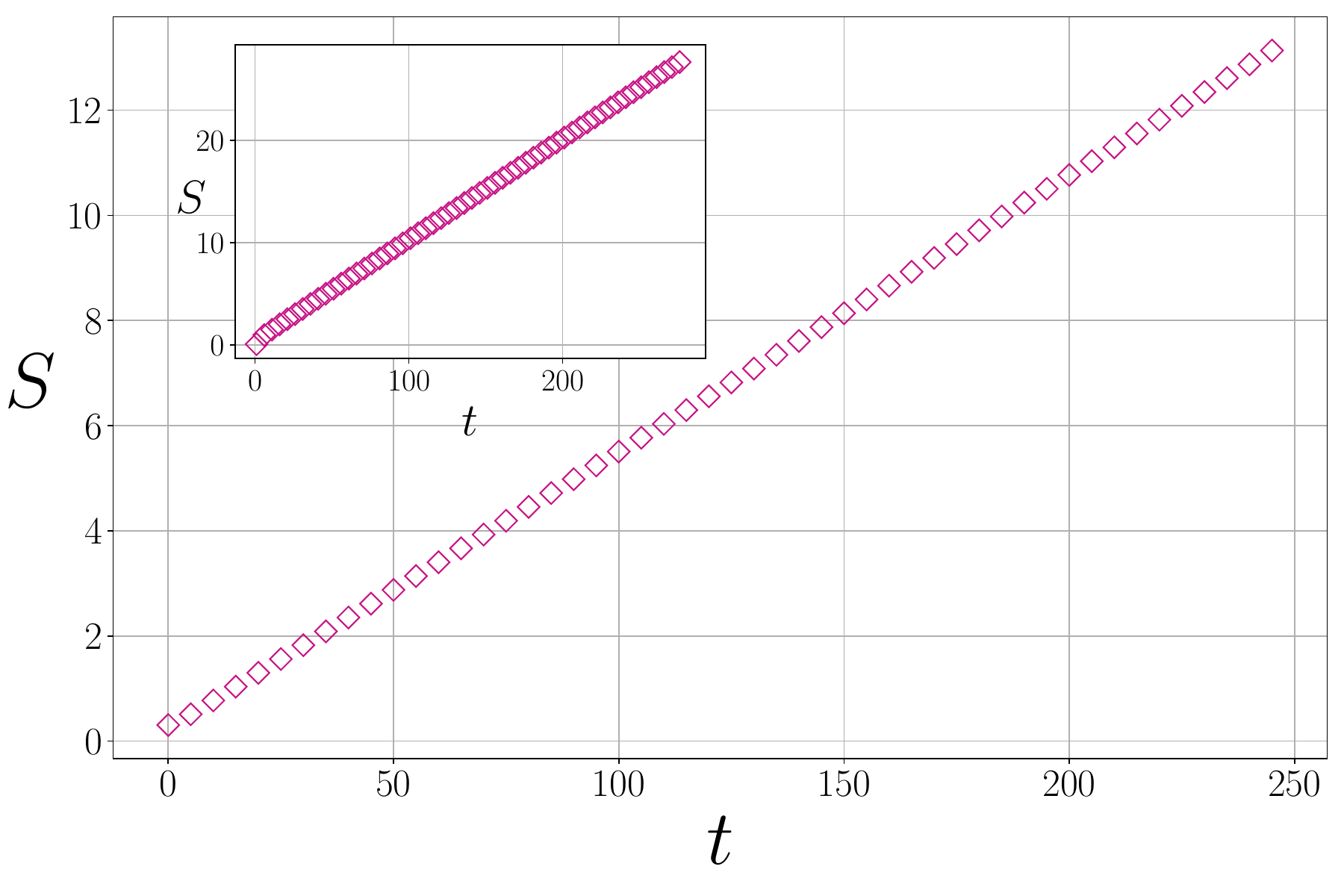}
    \caption{Growth of the entanglement entropy after a quench to an inhomogeneous Hamiltonian. 
    In the main plot we report the evolution of the half chain entanglement entropy after a quench from the ground state of the Ising chain with transverse field $h_0 = 3.0$ to the inhomogeneous Ising chain with magnetic field $h_L = 5.0$ on the left $j \leqslant0$ and $h_R = 4.0$ on the right $j>0$ in a system of total size $L = 1000$.
    We see that the entropy presents a clear linear growth.
    In the inset we instead report the entropy after a quench starting from the domain wall N\'eel state~\eqref{eq:dwneel}, equal to the N\'eel state for $j \leqslant 0$ and to the state with all spins down for $j>0$, to the inhomogeneous XX chain with left magnetic field $h_L = 0.0$ and right one $h_R = 1.0$.
    We can again observe a clear linear growth.
    }
    \label{fig:isingentropy}
\end{figure}

Let us focus on the situation $|h_L-h_R|\leqslant 2$, in which Eq.~\eqref{eq:twoside} predicts a linear entanglement growth with time. In Fig.~\ref{fig:isingentropy} we show numerical data for $S$ as a function of time. 
The data in the main figure are for the Ising chain. 
The chain is initially prepared in the ground state of the homogeneous Ising chain with $h_0=3$. Then, the system is let to evolve under the inhomogeneous Ising chan with $h_L=5$ and $h_R=4$.  The data are for a system with $L=10^3$ sites. In the inset of Fig.~\ref{fig:isingentropy} we provide data for the XX chain. The left part of the chain is initially prepared in the N\'eel state, whereas the right one in the state with all the spins down, which is the vacuum state for the fermions. Precisely, we have 
\begin{equation}\label{eq:dwneel}
    \ket{\psi_0} = \slr{\bigotimes_{j\leqslant 0} \ket{\uparrow}_{2j-1}\ket{\downarrow}_{2j} } \otimes \slr{\bigotimes_{j>0} \ket{\downarrow}_{2j-1}\ket{\downarrow}_{2j}}.
\end{equation}
At $t>0$ the system evolves with the inhomogeneous XX chain with $h_L=0$ and $h_R=1$.
The data in Fig.~\ref{fig:isingentropy} exhibit a clear linear growth with time for both models, as expected. Notice that at short times, deviations from the linear behavior are present.  
To obtain a more robust check of the results of Section~\ref{sec:xx-coeff} and~\ref{sec:ising-coeff} we fit the coefficient of the linear entropy growth.

\begin{figure}[t]
    \centering
    \includegraphics[width=0.48\linewidth]{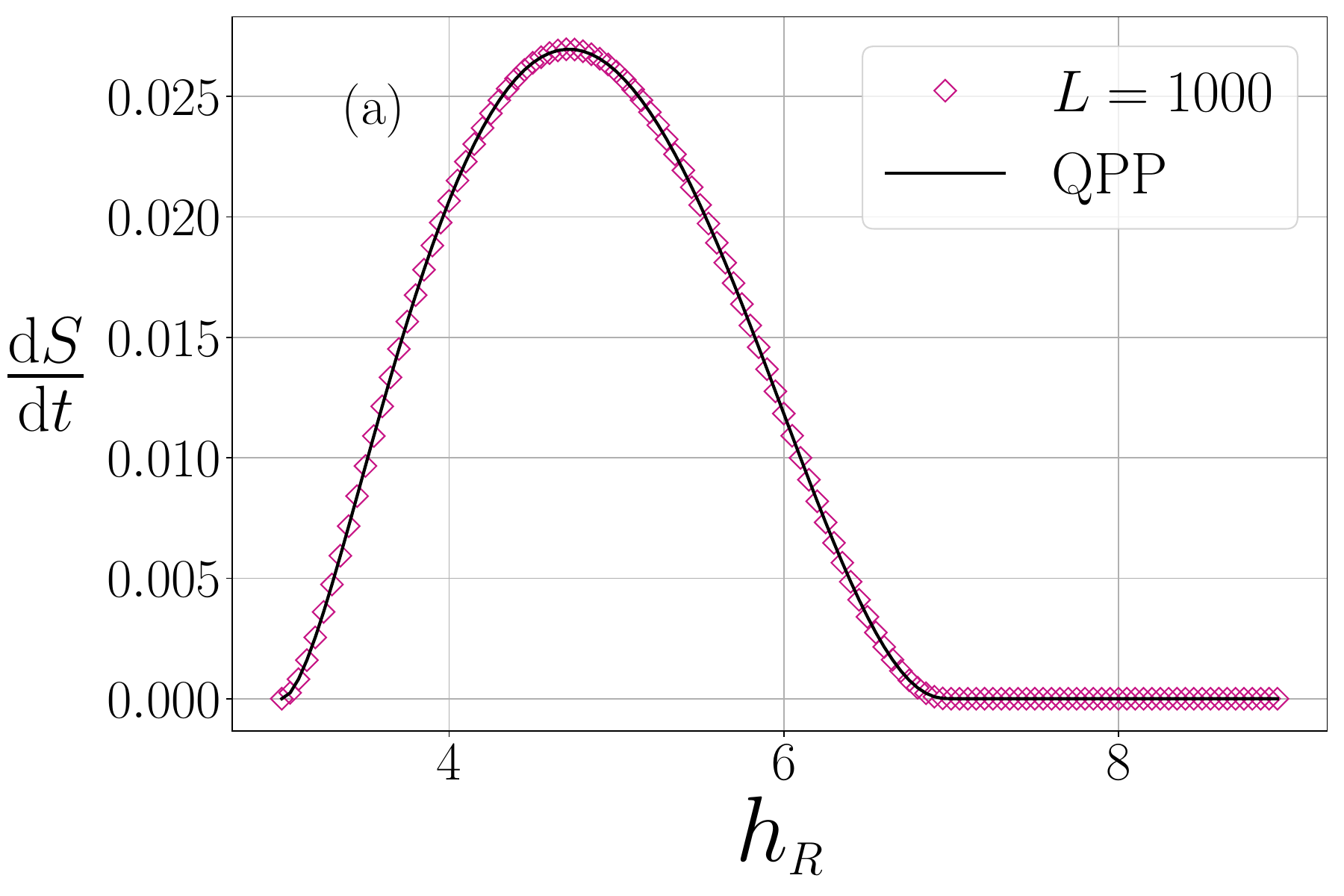}
    \includegraphics[width=0.48\linewidth]{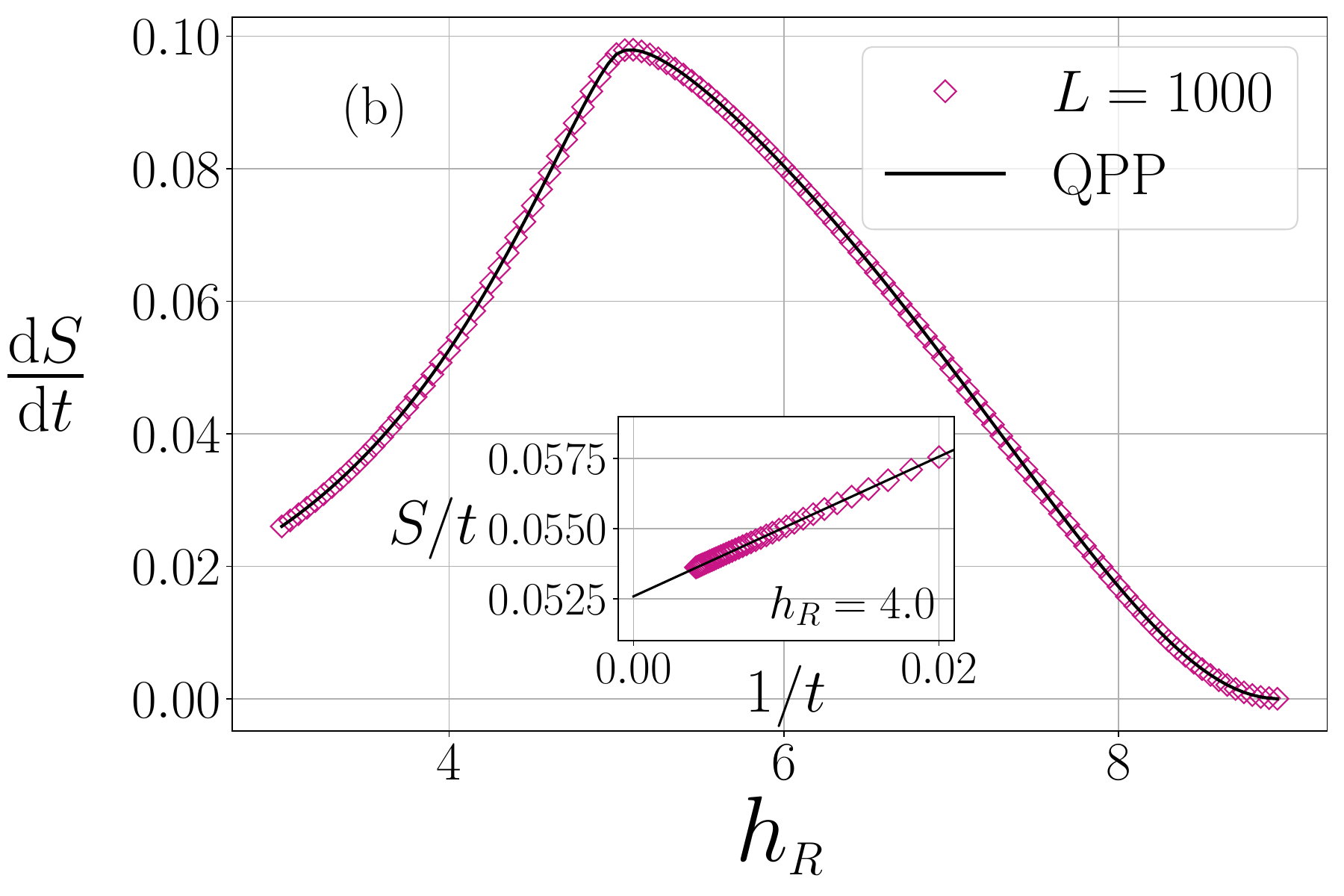}
    \caption{Slope of the linear growth of entanglement entropy after a quench from ground state of the Ising model with $h_0 = 3.0$.
    (a) Quench to inhomogeneous Ising model with $h_L = 3.0$, as a function of $h_R$. The symbols are the numerical results for the slopes obtained through a scaling analysis and the solid line is the theoretical prediction in \cref{eq:singleside}.
    (b) Quench to inhomogeneous Ising model with $h_L = 5.0$. Again the symbols are the numerical results and the solid line is the prediction in \cref{eq:twoside}.
    For both values of $h_L$ we observe a perfect agreement between the prediction and the numerics.
    In the inset of (b) we report the scaling analysis of the slope of the entropy for $h_L = 5.0$, $h_R = 4.0$.
    We plot $S/t$ as a function of $1/t$ and we extrapolate at $1/t \to 0$.
    As expected, for large times the leading correction to the linear growth of $S$ behaves as $1/t$. 
    }
    \label{fig:ising}
\end{figure}

\begin{figure}[t]
    \centering
    \includegraphics[width=0.7\linewidth]{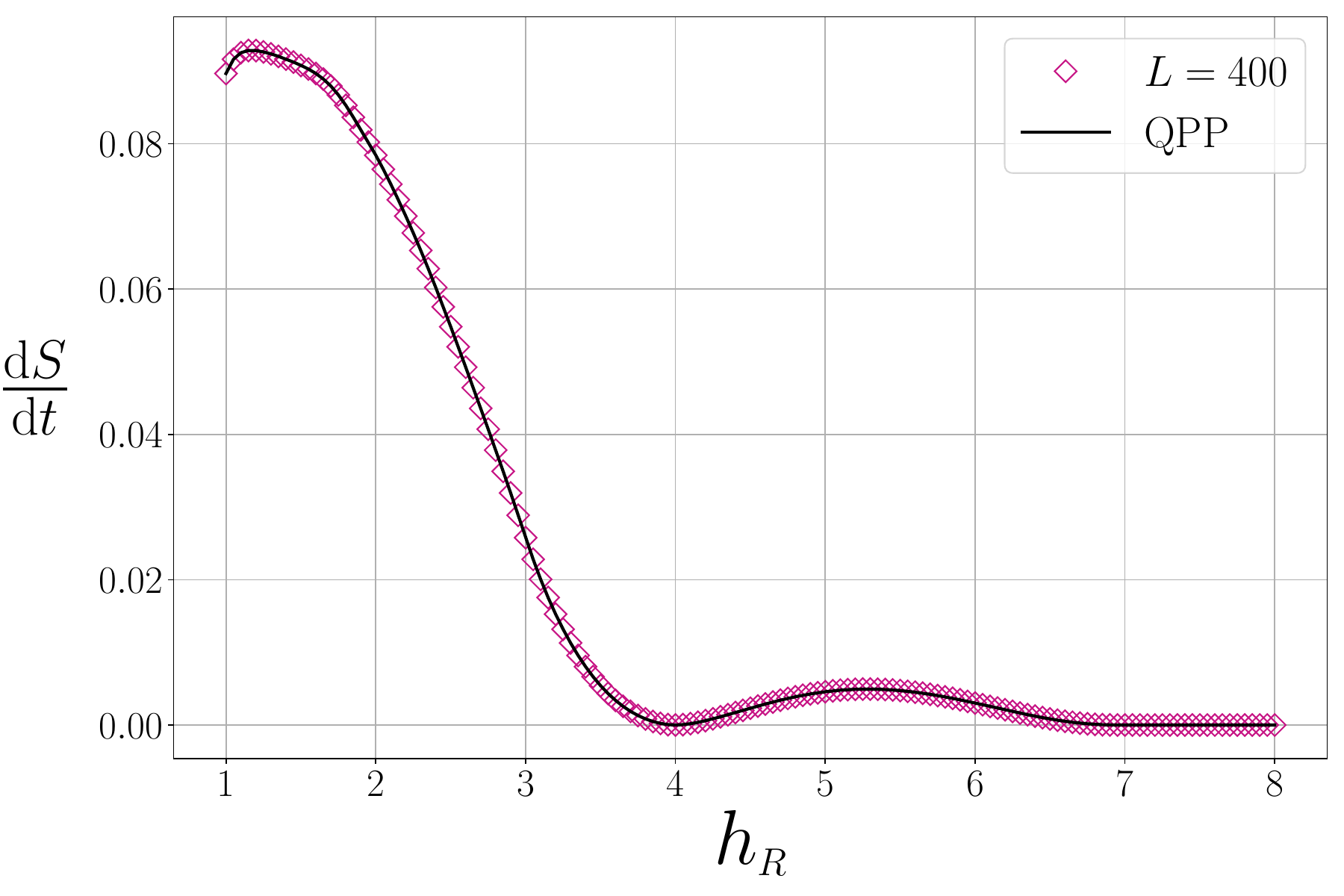}
    \caption{
    Slopes of the linear growth of the entanglement entropy after a quench from the ground state of the inhomogenous Ising model with $h_{0L} = 3.0$ and $h_{0R} = 4.0$, to the inhomogeneous Ising model with $h_{L} = 3.0$ as a function of $h_R$.
    The symbols are the numerical results obtained through a scaling analysis and the solid line is the prediction in \cref{eq:singleside}.
    We observe a perfect agreement.
    }
    \label{fig:isinginomog}
\end{figure}

Let us first focus on quenches in the inhomogeneous Ising chain. The prequench initial state is the same as in \cref{fig:isingentropy}, i.e., the ground state of the Ising chain at $t = 1$ and $h_0=3$. 
The system is then evolved with the Ising chain with $t = 1$, fixed $h_L$ and several values of $h_R$. 
In \cref{fig:ising}, we report the numerical results for the slopes of linear part of the entanglement entropy. The data are for $L=10^3$. 
In panel (a) and (b) we show results for $h_L=3$ and $h_L=5$, respectively. 
The symbols in the Figure are the numerically extracted slopes of the linear growth, whereas 
the solid black line is the prediction~\eqref{eq:singleside} obtained from the quasiparticle picture, were we  used the transmission coefficient (cf.~\eqref{eq:Tisi}). 
The agreement between the numerical data and the quasiparticle picture is perfect. Notice that at $h_R=3$ the slope of the entanglement growth vanishes. This happens because for $h_R = h_L =3$ the Hamiltonian is homogeneous and the initial state is its ground state, implying that the system does not evolve. 
On the other hand, the fact that $dS/dt$ is zero for $h_R>7$ is consistent with the condition 
that $|(h_L^2-h_R^2+4h_L\cos(k_L))/(4h_R)|> 1$ for any $k_L$ for transport to be forbidden. Notice that $dS/dt$ attains a maximum at $h_R\approx 5$. In \cref{fig:ising}(b) we show results for fixed $h_L=5$ as a function of $h_R$. In contrast with \cref{fig:ising} (a) 
now the dynamics is nontrivial for any value of $h_R$ because the initial state is never eigenstate of the post-quench Hamiltonian.
The qualitative behavior of the slope of the entanglement growth is the same as in \cref{fig:ising} (a). The entanglement growth is suppressed for $h_R>9$.  Finally, it is interesting to investigate finite-time corrections. In the inset of \cref{fig:ising} (b) we focus on the dynamics at $h_R=4$ plotting $S/t$ versus $1/t$. At long times the data exhibit linear behavior, suggesting that $S = a / t+b$ as expected.

In \cref{fig:isinginomog} we instead present the result for the slopes of the linear entanglement growth for a quench from an inhomogeneous state, the ground state of the inhomogeneous Ising chain with initial magnetic fields $h_{0L}=3.0$ on the left and  $h_{0R}=4.0$ on the right. 
The system is quenched to the inhomogeneous Ising chain with $h_L = 3.0$ and in \cref{fig:isinginomog} the slopes are plotted as a function of $h_R$.
The symbols in \cref{fig:isinginomog} are again the numerical result for the slopes, obtained through the same scaling analysis depicted in the inset of \cref{fig:ising} (b), while the solid black line is the quasiparticle prediction obtained from \cref{eq:singleside}. 
We see a perfect agreement between the prediction and the numerics. 
Notice in particular that the growth vanishes at $h_R = 4.0$, corresponding to the pre-quench inhomogeneous Hamiltonian, and for $h_R > 7$, similarly to \cref{fig:ising} (a).

\begin{figure}[t]
    \centering
    \includegraphics[width=0.7\linewidth]{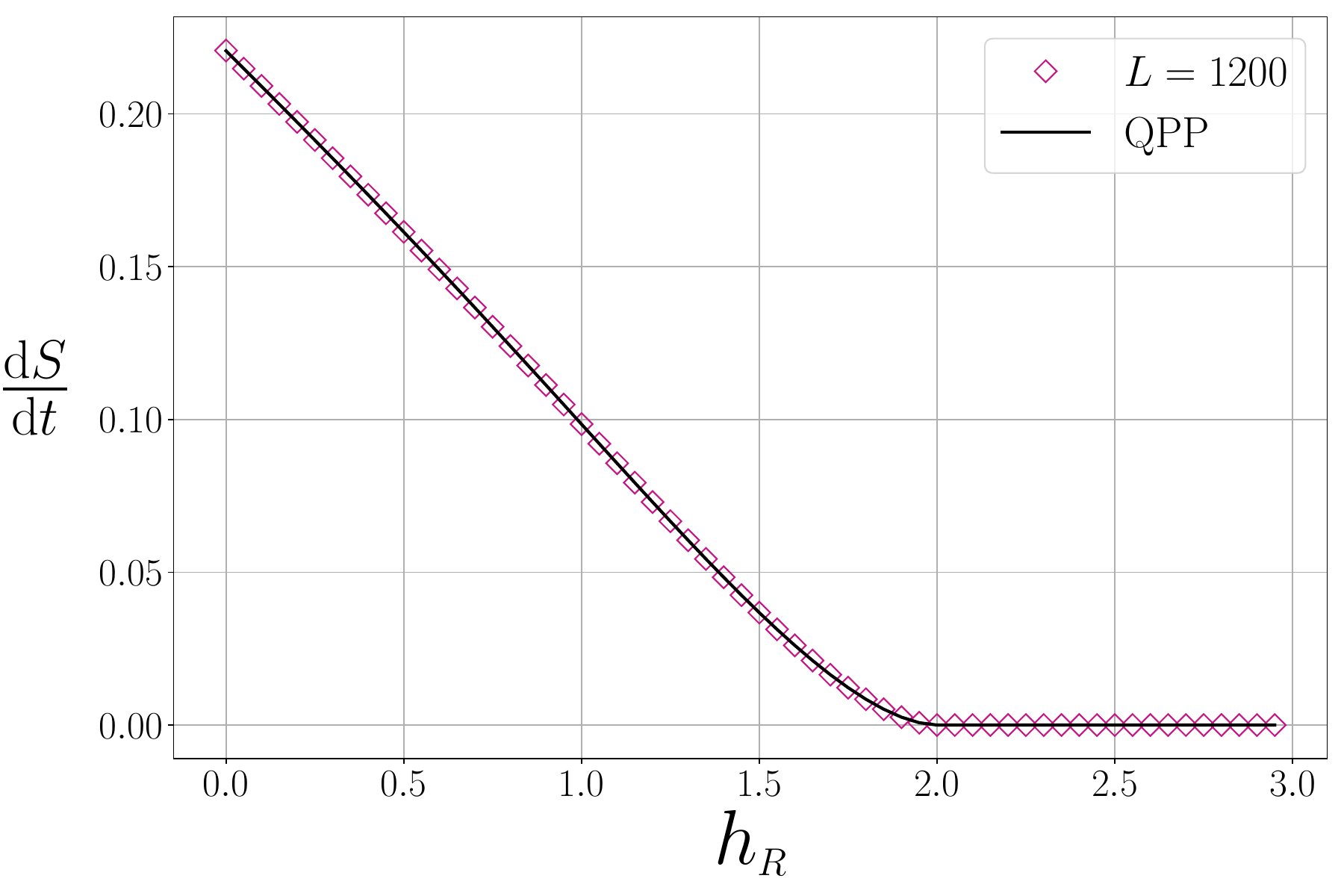}
    \caption{Slopes of the linear growth of the entanglement entropy after a quench from the state~\eqref{eq:dwneel} in the XX chain with $h_L=0$ as a function of $h_R$. The symbols are the numerical results for the slopes obtained through a scaling analysis and the solid line is the theoretical prediction in \cref{eq:singleside}. We observe a perfect agreement between the prediction and the numerics.}
    \label{fig:XX_chain}
\end{figure}
\begin{figure}[t]
    \centering
    \includegraphics[width=0.7\linewidth]{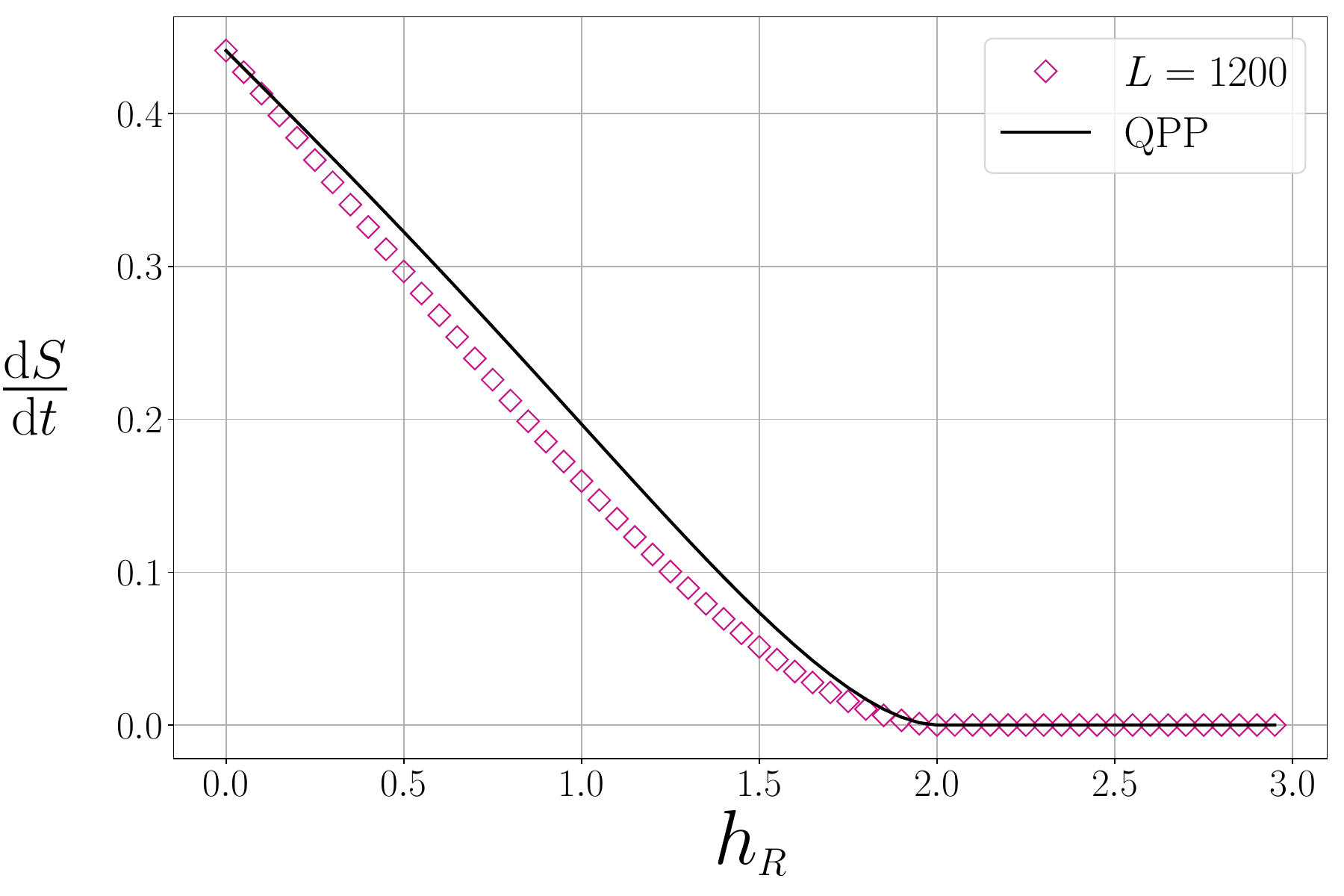}
    \caption{Slopes of the linear growth of the entanglement entropy after a quench from the N\'eel state in the XX chain with $h_L=0$ as a function of $h_R$. 
    The symbols are the numerical results for the slopes and the solid line is the theoretical prediction in \cref{eq:twoside}. 
    We observe a deviation from the prediction, signalling a failure of \cref{eq:twoside} in the XX chain for a quench on both sides.
    }
    \label{fig:neel_XX_chain}
\end{figure}

In \cref{fig:XX_chain} we discuss the slope of the linear entanglement growth after quenches in the XX chain. The system is prepared in the initial state~\eqref{eq:dwneel}. 
At time $t >0$ we perform a sudden quench to the inhomogeneous XX chain with $J = 1$, $h_L=0$ and varying $h_R$. 
Since the state with all the spins down is an eigenstate of the XX chain, while the N\'eel state is not, the dynamics is nontrivial only in the left part of the chain. The symbols in Fig.~\ref{fig:XX_chain} are the numerical results for $dS/dt$, whereas the black line is the quasiparticle prediction (cf.~\eqref{eq:singleside}) where we used~\eqref{eq:xxT} for the transmission coefficient. The agreement between the quasiparticle picture and the numerics is perfect. At $h_R=0$ the rate of the entanglement growth has a maximum. This corresponds to the situation in which the Hamiltonian is homogeneous. At $h_R>0$ the rate decreases monotonically up to $h_R=2$, where it vanishes. This is consistent with the condition $|h_R-h_L|\leqslant 2$ for transport and entanglement growth to be allowed. 

On the other hand, \cref{fig:neel_XX_chain} depicts the slopes of the entanglement entropy growth after a quench in the inhomogeneous XX chain with $J= 1, h_L = 0$ starting from the N\'eel state. 
In this case, the quench happens on both sides of the chain, rather than only in the left one as in \cref{fig:XX_chain}.
Now we see that quasiparticle prediction~\eqref{eq:twoside} (solid black line) is not in agreement with the result of the numeric computation (symbols), except at $h_R = 0$, i.e., in the homogeneous quench, and for $|h_R-h_L|\leqslant 2$ at which entanglement growth is suppressed. 
A potential reason for the this discrepancy is the presence of coherent cancellation of correlations, which is possible in this case because the fermionic modes are the same in the two parts of the chains since the eigenstates of the XX chain do not depend on the magnetic field, differently from the Ising chain.
We leave the analytic study of this quench for future work.

\subsection{Entanglement growth in the XXZ chain}
\label{sec:xxz}

\begin{figure}[t]
    \centering
    \includegraphics[width=0.75\linewidth]{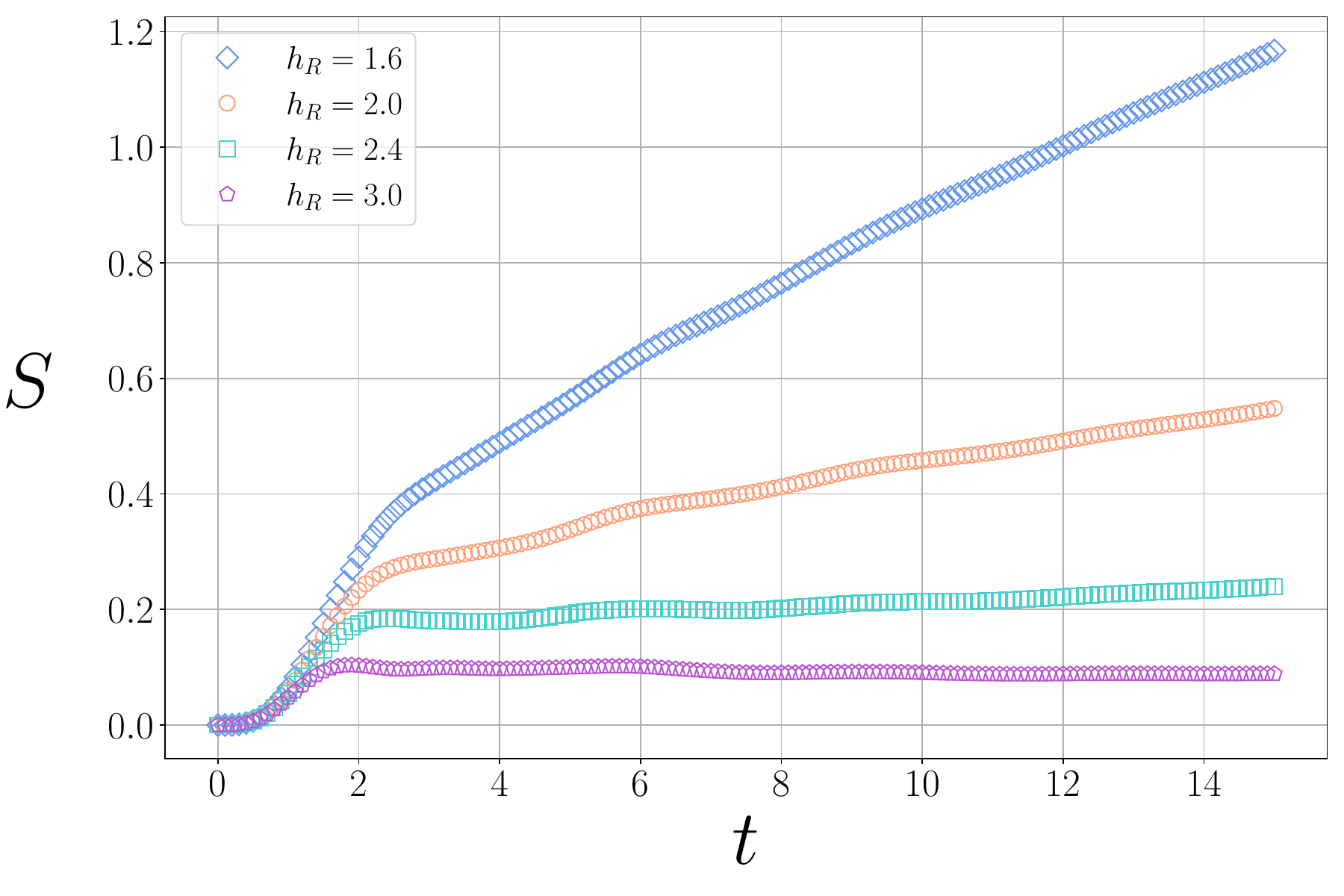}
    \caption{Entanglement entropy as a function of time for a quench from the domain wall N\'eel state~\eqref{eq:dwneel} to the inhomogeneous XXZ chain with $\Delta = 0.7$ and magnetic field on the left $h_L=0$. 
    The symbols are the numerical results for the entropies for different values of the magnetic fields on the right $h_R = 1.6, 2.0, 2.4,$ and $3.0$.
    The simulations have been performed with TEBD algorithm with maximum bond dimension $\chi = 800$ for all values of $h_R$.
    Comparing with \cref{fig:XX_chain}, we observe that while in the XX chain the entanglement entropy always saturates for $h_R \geqslant 2$, at $\Delta = 0.7$ the entropy presents a small linear growth even at $h_R = 2.4$, while it saturates for $h_R = 3.0$.
    }
    \label{fig:XXZ_entropy_0.7}
\end{figure}

\begin{figure}[t]
    \centering
    \includegraphics[width=0.75\linewidth]{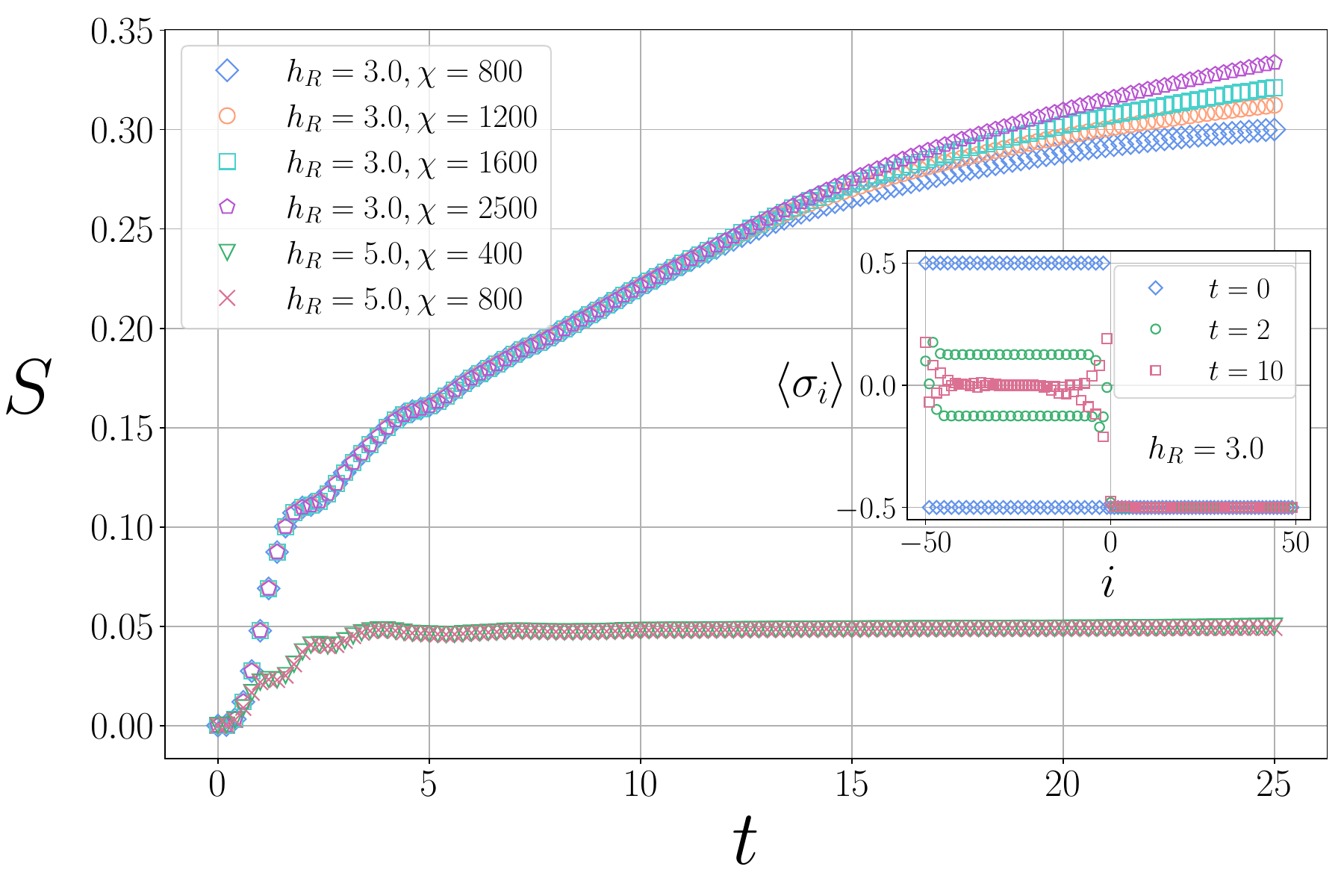}
    \caption{Entanglement entropy as a function of time for a quench from the domain wall N\'eel state to the inhomogeneous XXZ chain with $\Delta = 2.0$ and magnetic field on the left $h_L=0$. The symbols are the numerical results for the entropies for two values of the magnetic field on the right $h_R = 3.0$ and $5.0$
    The simulations have been performed with TEBD algorithm with different values of the maximum bond dimension $\chi$ depending on $h_R$.
    We see that at $\Delta=2.0$ even at magnetic field $h_R = 3.0$ the entropy presents a linear growth. 
    For times $t \lesssim 15$, the algorithm has already converged at $\chi = 800$ and the entropies presents a clear linear growth; while for $t \gtrsim 15$ instead, the entropy is still growing with $\chi$ even at $\chi = 2500$, compatibly with a continuation of the linear growth in time.
    For $h_R = 5.0$, instead, the entropy saturates and the algorithm has reached convergence already at $\chi = 400$.
    In the inset, we report the value of the local magnetisation $\avg{\sigma^z_i}$ as a function of the lattice site $i$ for the magnetic field $h_R = 3.0$ at times $t = 0, 2,$ and $10$.
    At time $t = 0$, the state is the domain wall N\'eel state with a N\'eel state for $i<0$ and with all down spins for $i>0$.
    At later times we see that while the magnetisation presents a non-trivial evolution for $i < 0$, it remains constant for $i > 0$, indicating the absence of transport between the two halves.
    Comparing with the main plot, we see that the systems displays linear growth of entanglement at $h_R = 3.0$ even in the absence of transport.
    }
    \label{fig:XXZ_entropy_2}
\end{figure}

Finally, we study numerically the behaviour of the entanglement entropy after a quantum quench to an inhomogeneous \emph{interacting} model. Here we consider the inhomogeneous XXZ chain~\eqref{eq:xxz-ham}. 
As for the XX chain, we initialise the system in the state~\eqref{eq:dwneel}, and at time $t>0$ we let it evolve unitarily with \cref{eq:xxz-ham} with fixed $h_L = 0$ and several values of $h_R$. We consider the XXZ chain with $\Delta = 0.7$ and $\Delta=2$. 

As for the noninteracting case, the dynamics in nontrivial only on the left side, since the state with all the down spins is an eigenstate of the XXZ chain. 
Since the XXZ chain is interacting, simulating the out-of-equilibrium dynamics is a challenging task. Here we employ the Time Evolving Block Decimation (TEBD) algorithm~\cite{schollwoeck2011the,paeckel2019time}. 
The initial state admits a straightforward representation as a Matrix Product State (MPS) with bond dimension $\chi=1$. The dynamics is generated by applying a fourth order trotterisation of the evolution operator. We verified that a Trotter step $\delta t=0.1$  is sufficient to obtain accurate results. After each evolution step the bond dimension increases.  To prevent the bond dimension from getting too large, we compress the MPS by performing a Singular Value Decomposition (SVD) and keeping only the largest $\chi_{max} = 2500$ singular values. 

In \cref{fig:XXZ_entropy_0.7} we show data for  $S$ as a function of time for a quench to the inhomogeneous XXZ chain with $\Delta = 0.7$, left magnetic field $h_L = 0$ and different values of the right field $h_R = 1.6, 2.0, 2.4,$ and $3.0$. 
We consider a system of size $L = 40$ and we perform the time evolution up to time $t = 15$, with a maximum bond dimension $\chi = 800$ for all values of $h_R$. 
By comparing the data with those obtained with $\chi = 400$ (not shown) we verified that $\chi = 800$ is sufficient to ensure convergence of the TEBD for all the values of $h_R$. While for $\Delta=0$, i.e., for  the XX chain  (see \cref{fig:XX_chain}), the linear growth of the entanglement entropy is absent for  all $h_R\geqslant 2.0$, for $\Delta=0.7$ the entropy exhibits a  linear growth with time  at $h_R = 2$ and $h_R=2.4$, 
whereas it saturates at $h_R = 3.0$. This is surprising because the XXZ chain is integrable, and the single-particle dispersion has the same form as for the XX chain~\cite{takahashibook}. This implies that for $|h_L-h_R|> 2$ energy conservation across the interface should prevent transport and hence entanglement growth. 
This surprising behavior is observed also at $\Delta>1$. 
In \cref{fig:XXZ_entropy_2}, we show data for $\Delta = 2.0$ and right magnetic field $h_R = 3.0$ and $5.0$ for a chain with $L = 100$,  times up to $t = 25$. For $h_R=3$ the different symbols are numerical results with different bond dimension $\chi$. Clearly, the TEBD data are reliable up to $t\lesssim 15$. At larger times it is difficult to conclude whether the extrapolation to $\chi\to\infty$ would confirm the linear growth of the entanglement entropy.  
Clearly,  the linear entanglement growth  persists even at $h_R = 3.0$ up to $t\lesssim 15$. 
For $h_R = 5.0$ the TEBD are converged already for $\chi = 800$, and the entropy saturates already at $t\approx 5$. It is interesting to clarify the relationship between entanglement growth and transport of local conserved quantities. In the inset of \cref{fig:XXZ_entropy_2}, we plot the local magnetisation $\avg{\sigma_i^z}$ as a function of the site $i$ and for several times.  At $t=0$ one has $\langle \sigma_i^z\rangle=(-1)^i/2$ for $i\leqslant 0$, and zero for $i>0$.  At later times the initial staggering gets  smoothened by the dynamics, and eventually the systems reaches a zero-magnetization profile. Crucially, while the magnetisation profile evolves  non-trivially for $i \leqslant 0$, there is no magnetisation transport from the left to the right subsystem, even at a sub-ballistic level. 
Additionally, we mention that  we verified that energy transport across the interface is also suppressed.
These results are remarkable, as they imply that the entropy growth happens despite the absence of transport of local charges.
This suggests that the growth of the entanglement entropy for $|h_L-h_R|$ could be attributed to transport of higher local conserved quantities, or to quasilocal ones.

\section{Conclusions}
\label{sec:concl}

We have investigated entanglement dynamics under inhomogeneous Hamiltonians $H = H_L + H_R$, where $H_L$ and $H_R$ are homogeneous Hamiltonians acting nontrivially only on the left and right parts of the system, respectively. We focused on the XX chain and the transverse-field Ising chain, with different magnetic fields in the two parts. Both models can be mapped to free-fermion systems. We conjectured an analytical formula for the dynamics of the entanglement entropy between the left and right regions. The prediction is based on the quasiparticle picture for entanglement spreading and holds in the hydrodynamic limit of asymptotically long times. Similar to the case of dynamics in the presence of localized defects, quasiparticles produced in the two regions undergo scattering when crossing the interface, generating entanglement between the reflected and transmitted quasiparticles. The entanglement content of the quasiparticles is expressed in terms of the transmission probability across the interface, which is obtained by solving a single-particle lattice Schr\"odinger equation for the stationary state.

Several interesting directions remain for future work. First, as anticipated, for the XX chain the quasiparticle picture fails to capture the entanglement dynamics for initial states with nonzero fermionic occupation in both parts of the chain. Clarifying the reason for this requires an \textit{ab initio} derivation of the quasiparticle picture. For the XX chain this is challenging but feasible, for instance by using the methods of Refs.~\cite{alba2022noninteracting,alba2021unbounded,caceffo2023entanglement,alba2023logarithmic}. Furthermore, as noted in Section~\ref{sec:main}, scattering at the interface is expected to generate entanglement among more than two quasiparticles. This could be detected using the tripartite information~\cite{caceffo2023negative,carollo2022entangled}. Our results suggest that in the presence of interactions, the linear growth of the entanglement entropy persists even in parameter regions where magnetization transport is forbidden. It would be interesting to clarify whether this is a finite-time effect or survives in the hydrodynamic limit. In the latter case, it would be interesting to understand the relationship with transport. Moreover, it should be possible to employ the framework of generalized hydrodynamics~\cite{bertini-2016,olalla-2016,bastianello2019generalized} to describe dynamics under the inhomogeneous XXZ chain Hamiltonian. Another avenue is to investigate entanglement dynamics in the presence of integrability-breaking interactions. Finally, it would be interesting to derive the dynamics of the logarithmic negativity~\cite{gruber2020time,caceffo2023entanglement,fraenkel2021entanglement,fraenkel2023extensive}.

\section*{Acknowledgements}

This study was carried out within the National Centre on HPC, Big Data and Quantum Computing - SPOKE 10 (Quantum Computing) and received funding from the European Union NextGenerationEU - National Recovery and Resilience Plan (NRRP) – MISSION 4 COMPONENT 2, INVESTMENT N. 1.4 – CUP N. I53C22000690001. 
VA and FR have been supported by the project ``Artificially devised many-body quantum dynamics in low dimensions - ManyQLowD'' funded by the MIUR Progetti di Ricerca di Rilevante Interesse Nazionale (PRIN) Bando 2022 - grant 2022R35ZBF. This article is based upon work from COST Action "Many-body Open Quantum Systems" (QOpen) CA24109 supported by COST (European Cooperation in Science and Technology).

\bibliographystyle{ytphys}
\bibliography{bibliography}
\end{document}